\documentclass[12pt]{article}
\usepackage{graphicx}
\usepackage{color}
\usepackage{cancel}
\usepackage{amsmath}
\usepackage{amsfonts}
\usepackage{amssymb}
\usepackage{hyperref}
\newcommand{\be}{\begin{equation}}
\newcommand{\ee}{\end{equation}}

\definecolor{redcom}{rgb}{1,0.1,0.2}
\definecolor{querycol}{rgb}{0.2,0.2,1}

\definecolor{purplerep}{rgb}{1,0.1,1}

\definecolor{green}{rgb}{0.4,0.8,0.4}

\begin{document}
\title{Sub-Compton Quantum Non-Equilibrium \\and Majorana Systems}
\date{}
\author{Samuel Colin\footnote{
Department of Physics and Astronomy, Clemson University, 120-A Kinard Laboratory, Clemson,\newline SC 29631-0978, USA.\newline
Centre for Quantum Dynamics, Griffith University, 170 Kessels Road, Brisbane, QLD 4111, Australia.\newline
\texttt{Email: scolin@clemson.edu}}
}
\maketitle
\vglue -1.8truecm

\abstract{We study the Majorana equation from the point of view of the de Broglie-Bohm pilot-wave theory (according to which a quantum ensemble of fermions is not only 
described by  a spinor but also by a distribution of position configurations). 
Although the \mbox{Majorana} equation involves a mass parameter, we show that the positions undergo luminal motion. 
In the case of free systems, we also show that the trajectory can be strongly helical  
(the diameter of the helix being the \mbox{Compton} wavelength). 
On a coarse-grained level (coarse-graining with respect to the \mbox{Compton} wavelength), these trajectories appear subluminal. 
The peculiar nature of the Majorana trajectory suggests a study of the temporal evolution of quantum non-equilibrium distributions, 
which are distributions allowed in the pilot-wave theory, in which the configurations are not distributed according to Born's law. 
We do such simulations for Dirac and \mbox{Majorana} systems, and we investigate whether quantum 
non-equilibrium might survive at the sub-Compton scale in systems described by Majorana spinors.}

\smallskip
\smallskip
{\bf Keywords:} de Broglie-Bohm pilot-wave theory, Dirac equation, Majorana equation, quantum non-equilibrium, relaxation to quantum equilibrium.

\section{Introduction}
The idea that a neutral fermion might be its own particle was shown to be theoretically possible by Majorana in \cite{majorana}, where it was already suggested 
that the neutrino might be such a particle. 
Since then, this concept has been found relevant in many different areas of physics. 
For example, in the last decades, Majorana spinors have been used in supersymmetric models, the reason being that, 
like Weyl spinors, they have the same number of degrees of freedom as real vector fields and can be superpartners to neutral bosons. 
Experiments are still running today, aiming to reveal the true nature of the neutrino, 
one of the signatures of a Majorana neutrino being  the neutrinoless double beta decay $2n\rightarrow 2p+2e^{-}$. 
 Although no `real' Majorana particle has ever been observed, 
it has been proposed that the Majorana equation could be simulated experimentally, 
thanks to an analog ion-trap experiment \cite{gerritsma} (in the same manner as the Dirac equation was simulated \cite{gerritsma0}), and very recently 
it was reported that  Majorana modes had been detected in an experiment involving a hybrid semiconductor-superconductor device \cite{kouwenhoven}. 
The wide relevance of the Majorana equation in these fields (high-energy physics, condensed matter and quantum information) was described in \cite{wilczek}. 

In this article, the Majorana equation will be looked at from the point of view of the de Broglie-Bohm pilot-wave theory - which is a theory first proposed by de Broglie 
\cite{debroglie28} and later rediscovered by Bohm \cite{bohm521,bohm522}, albeit in a slightly different form -  
according to which fermions are not only described by spinors but also by actual position configurations. 
The de Broglie-Bohm theory has been extensively developed but mostly for Dirac spinors when it comes to relativistic fermions.
According to this theory (of which a presentation can be found in \cite{holland}), a single Dirac particle is not only described by its four-component spinor $\psi_D(t,\vec{x})$ but also by its position $\vec{x}_D(t)$.
A quantum ensemble is then described by a spinor and by a distribution of particle positions, that we denote by $\rho_D(t,\vec{x})$. 
In a quantum ensemble of Dirac particles, described by the same spinor $\psi_D(t,\vec{x})$, the position configurations will be distributed according 
to the standard probability density for all times 
\be\label{queq}
\rho_D(t,\vec{x})=\psi^\dagger_D(t,\vec{x})\psi_D(t,\vec{x})
\ee
if it happens to be true for some initial time. This ensures that the predictions of the standard interpretation are reproduced and it is  a 
consequence of the fact that the Dirac particles move according to the guidance equation
\be
\vec{v}_D(t)=\frac{\vec{j}_D(t,\vec{x})}{j^0_D(t,\vec{x})}\bigg{|}_{\vec{x}=\vec{x}(t)}~,
\ee
where $\vec{v}_D$ is the velocity and $j^\mu_D=(j^0_D,\vec{j}_D)$ is the Dirac current satisfying the continuity equation. 

Ensembles such as (\ref{queq}) are said to be in quantum equilibrium  
and they only form a small class among all the ensembles that can be considered in the de Broglie-Bohm pilot-wave theory.
Indeed, in principle, it is possible to consider ensembles in which the configurations are not distributed according to Born's law initially:
they are referred to as ensembles in quantum non-equilibrium \cite{valentini92}.  
Experiments performed on such systems would violate the predictions of standard quantum mechanics \cite{valentini02}. 
If this possibility is considered seriously, one has to explain why we don't see these ensembles today, if they existed at some point in the history of the universe.
This is done by invoking the idea of relaxation to quantum non-equilibrium \cite{valentini91a,valentini91b,valentini92}, which in essence says that 
a non-equilibrium distribution is quickly driven to quantum equilibrium thanks to the guidance equation, 
provided that the system has enough complexity or that it gets entangled. 
This does not preclude that more exotic systems would still be in quantum non-equilibrium today \cite{valentini07} or that the existence of quantum non-equilibrium in the early 
universe could leave other imprints, observable today, for example in the cosmic microwave background \cite{valentini08}. 
Relaxation is  supported by various numerical simulations \cite{valentini042,cost10,toruva,colin2012}.

We shall see that the pilot-wave theory for the Majorana equation is very similar, in its construction, to the pilot-wave theory for the Dirac equation.
However a Majorana spinor satisfies an additional constraint, namely it is invariant under charge conjugation. 
This implies, as we shall demonstrate, that the motion of the position configuration is luminal, despite the fact that the mass appears in the Majorana equation. 
We will also show that the trajectory can be helical, the diameter of the helix being the Compton wavelength. 
On a coarse-grained level, at which 
details about the Compton scale are being erased, the Majorana trajectory appears subluminal. Due to the special nature of the trajectory, 
we shall investigate relaxation to quantum equilibrium for Majorana systems and in particular whether quantum non-equilibrium survives 
at scales below the Compton scale.

It is often stressed  that a classical Majorana spinor should not be confounded with the wave-function of an elementary particle such as the Majorana neutrino (this is discussed in \cite{mannheim}). Identifying the two concepts usually leads to strong claims and debates (see \cite{majorana1,majorana2,majorana3} for example). 
Indeed, although a classical Lagrangian can involve self-conjugate Majorana fields, once the theory is quantized, only 
the quantum field is self-conjugate (and the classical components associated to each annihilation or creation operator in the Fourier decomposition of the quantum field are not self-conjugate).
The same is true for the real Klein-Gordon quantum field theory: although the classical Klein-Gordon field is real, that does not mean that the amplitude 
associated to a neutral Klein-Gordon boson in the full-fledged quantum field theory is real. 
That being said, that last reasoning depends on which vacuum appears in the amplitude. The situation is also a bit different 
in a pilot-wave approach for the standard model of particle physics, 
when one realizes that all the particles are fundamentally massless and only acquire their bare mass when the Higgs field condenses \cite{cowi}.  
In that perspective, Weyl spinors seem to be the ultimate building blocks from which any Dirac spinor can be made of, but Majorana spinors 
can play the same role \cite{struyve_zz}. 
We will discuss this further in the conclusion.
Also, as far as we can see, Majorana spinors can be relevant for Majorana modes \cite{wilczek,wilczek2} because 
these modes are obtained by summing an annihilation and a creation operator acting on a Fermi sea (as such their effective wave-function are also self-conjugate). 

This article is organized in the following way. In section 2 we develop the pilot-wave theory for the Majorana equation. 
We show that the motion of the configuration is luminal and we illustrate the helical nature of the trajectory. 
In section 3 we identify systems appropriate for a study of relaxation to quantum equilibrium in Majorana systems. 
In sections 4 and 5 we present the results for such simulations, respectively in two and three dimensions, for both Dirac and Majorana systems.
In section 6 we investigate whether quantum non-equilibrium can survive at scales below the Compton wavelength.
In section 7 we conclude and we suggest a few ideas for further investigation.
\section{Pilot-wave theory for a Majorana spinor}
In this section, we construct the pilot-wave theory for a Majorana spinor in spacetimes of dimension $3+1$, $2+1$ and $1+1$.
We show that the motion is always luminal in $3+1$ and $2+1$ space times, cases for which we illustrate the helical nature of the trajectory.
\subsection{$3+1$ spacetime}
We start from the Dirac equation in a $3+1$ spacetime
\be\label{dirac}
(i\gamma^\mu\partial_\mu-m)\psi_D=0~
\ee
and use the Weyl representation for the $\gamma$-matrices
\be
\gamma^\mu=\begin{pmatrix}0 & \sigma^\mu\\ \tilde\sigma^\mu & 0\end{pmatrix}~,
\ee
where $\sigma^0=\tilde\sigma^0=1$ and $\tilde\sigma^j=-\sigma^j$.

A Majorana spinor $\psi_M$, by definition,  doesn't change under charge conjugation
\be\label{majorana}
\psi_M=C\psi^*_M=i\gamma^2\psi^*_M~.
\ee
To build such a solution, we add a usual solution of the Dirac equation and its charge conjugate:
\be\label{addcc}
\psi_M=\frac{1}{\sqrt{N}}(\psi_D+i\gamma^2\psi^*_D)~,
\ee
where $N$ is a normalization factor, not necessarily equal to 2 (note that $i\gamma^2\psi^*_D$ is also solution of (\ref{dirac})). 

If we write $\psi_M$ as $\begin{pmatrix}\psi_{M_1}\\\psi_{M_2}\end{pmatrix}$, (\ref{majorana}) implies that $\psi_{M_1}=i\sigma_2\psi^*_{M_2}$. Therefore the corresponding 4-current is 
\begin{align}
j^\mu_M=&\bar\psi_M\gamma^\mu\psi_M=\psi^\dagger_{M_2}\sigma^\mu\psi_{M_2}+\psi^\dagger_{M_1}\tilde\sigma^\mu\psi_{M_1}&\nonumber\\
=&\psi^\dagger_{M_2}\sigma^\mu\psi_{M_2}+\psi_{M_2}^T\sigma_2\tilde\sigma^\mu\sigma_2\psi^*_{M_2}
=2\psi^\dagger_{M_2}\sigma^\mu\psi_{M_2}~.&
\end{align}
The last line follows from the fact that $\sigma_2$ is purely imaginary, anti-commutes with $\sigma_1$ and $\sigma_3$ and its square is the identity.
Finally any current $\Phi\sigma^\mu\Phi$ is light-like in a $3+1$ spacetime (this is seen after easy algebraic manipulations). 
Therefore the Majorana current $j^\mu_M$ is light-like.\footnote{In a $2+1$ spacetime, the current $\Phi\sigma^\mu\Phi$ is time-like. Therefore the motion 
of a Dirac particle in a $2+1$ spacetime, when one uses a two-dimensional representation of the Clifford algebra, is subluminal.}

In the corresponding pilot-wave theory, an element of a Majorana system is described by its 4-spinor $\psi_M(t,\vec{x})$ together with its position $\vec{x}_M(t)$. 
The position configuration evolves according to the guidance equation
\be
\vec{v}_M(t)=\frac{\vec{j}_M(t,\vec{x})}{j^0_M(t,\vec{x})}\bigg|_{\vec{x}=\vec{x}_M(t)}~.
\ee
Because the Majorana current is light-like, the motion is always luminal.
\subsubsection{Example 1: plane-wave solution}
We consider the following right-handed Dirac spinor
\be\label{freedirac}
\psi_D(t,\vec{x})=\begin{pmatrix}\sqrt{\frac{E-p}{2E}}\\0\\ \sqrt{\frac{E+p}{2E}}\\0\end{pmatrix}
e^{-iEt+ip_z z}
\ee
whose associated 4-current is 
\begin{eqnarray} 
j^\mu_D=\bar\psi_D\gamma^\mu\psi=(1,0,0,\frac{p_z}{E})~.
\end{eqnarray}
Therefore, in the corresponding pilot-wave theory, 
the Dirac configuration moves along a straight line with uniform velocity $(0,0,\frac{p_z}{E})$.

\begin{figure}
\includegraphics[width=\textwidth]{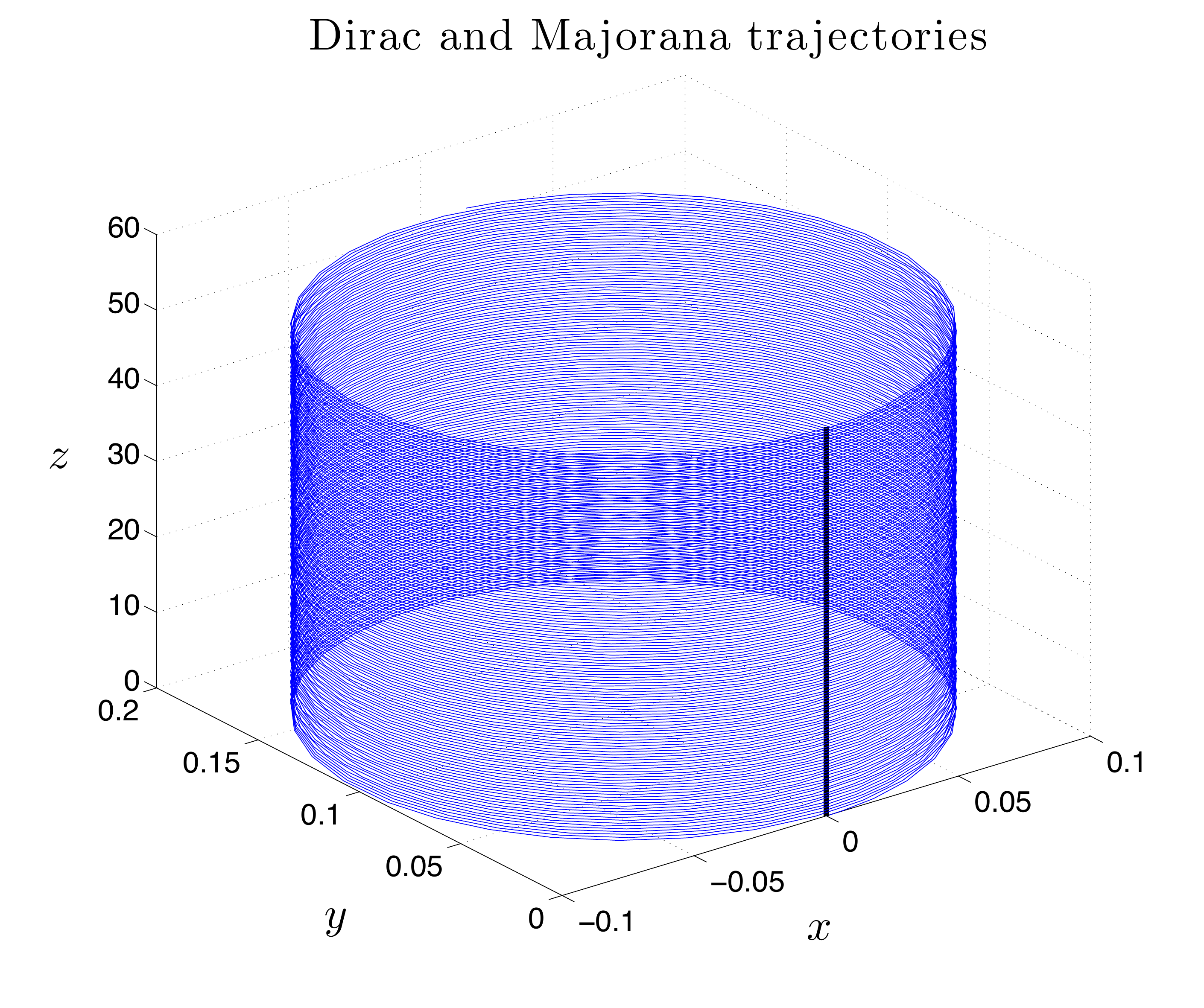}
\caption{\label{fig1}Both configurations start from the origin at $t=0$.
The configuration guided by the Dirac (resp. Majorana) spinor follows the black (resp. blue) trajectory during the time-interval 
$[0,100].$The Dirac spinor is the one given in (\ref{freedirac}) with $m=5$ and $p_z=3$. 
The Majorana spinor is obtained from the Dirac spinor thanks to (\ref{freemajo}).}
\end{figure}

If we consider a Majorana solution instead, then the spinor is given by 
\be\label{freemajo}
\psi_M=\frac{1}{\sqrt{N}}(\psi_D+i\gamma^2\psi^*_D)=\frac{1}{\sqrt{N}}(\psi_D+\psi_{D_c})~
\ee
and the motion is not uniform anymore. Indeed, the current reads
\be
\bar\psi_M\gamma^\mu\psi_M=\frac{2}{N}(j^\mu_D+\mathfrak{Re}(\bar\psi_{D_c}\gamma^\mu\psi_D))=\frac{2}{N}(j^\mu_D+j^\mu_{D_{c}-D})~,
\ee
and in the case of (\ref{freedirac}), we have that
\be
j^\mu_{D_{c}-D}=(0,\frac{m}{E}\cos{(2Et-2p_z z)},-\frac{m}{E}\sin{(2Et-2p_z z)},0)~,
\ee
so that the total current $j^\mu_M$ is given by 
\be
(1,\frac{m}{E}\cos{(2Et-2p_z z)},-\frac{m}{E}\sin{(2Et-2p_z z)},\frac{p}{E})~.
\ee
Therefore the velocity field is 
\be
v^k_M=\frac{\bar\psi_M\gamma^{k}\psi_M}{\psi^\dagger_M\psi_M}=(\frac{m}{E}\cos{(2Et-2p_z z)},-\frac{m}{E}\sin{(2Et-2p_z z)},\frac{p}{E})~,\quad
\ee
whose solution is 
\be
\vec{x}_M(t)=(\frac{1}{m}\sin{(2\frac{m^2}{E}t)},\frac{1}{m}\cos{(2\frac{m^2}{E}t)},\frac{p}{E}t)
\ee
if $\vec{x}_M(t=0)=\vec{0}$. This is the equation of a helical trajectory whose diameter is the Compton wavelength  (an example is illustrated 
in Figure \ref{fig1}).
In this special case, we therefore understand how the configuration, despite undergoing luminal motion, appears to move subluminally on a coarse-grained level.
\subsubsection{Example 2: superposition of plane-wave solutions}
\begin{figure}
\includegraphics[width=\textwidth]{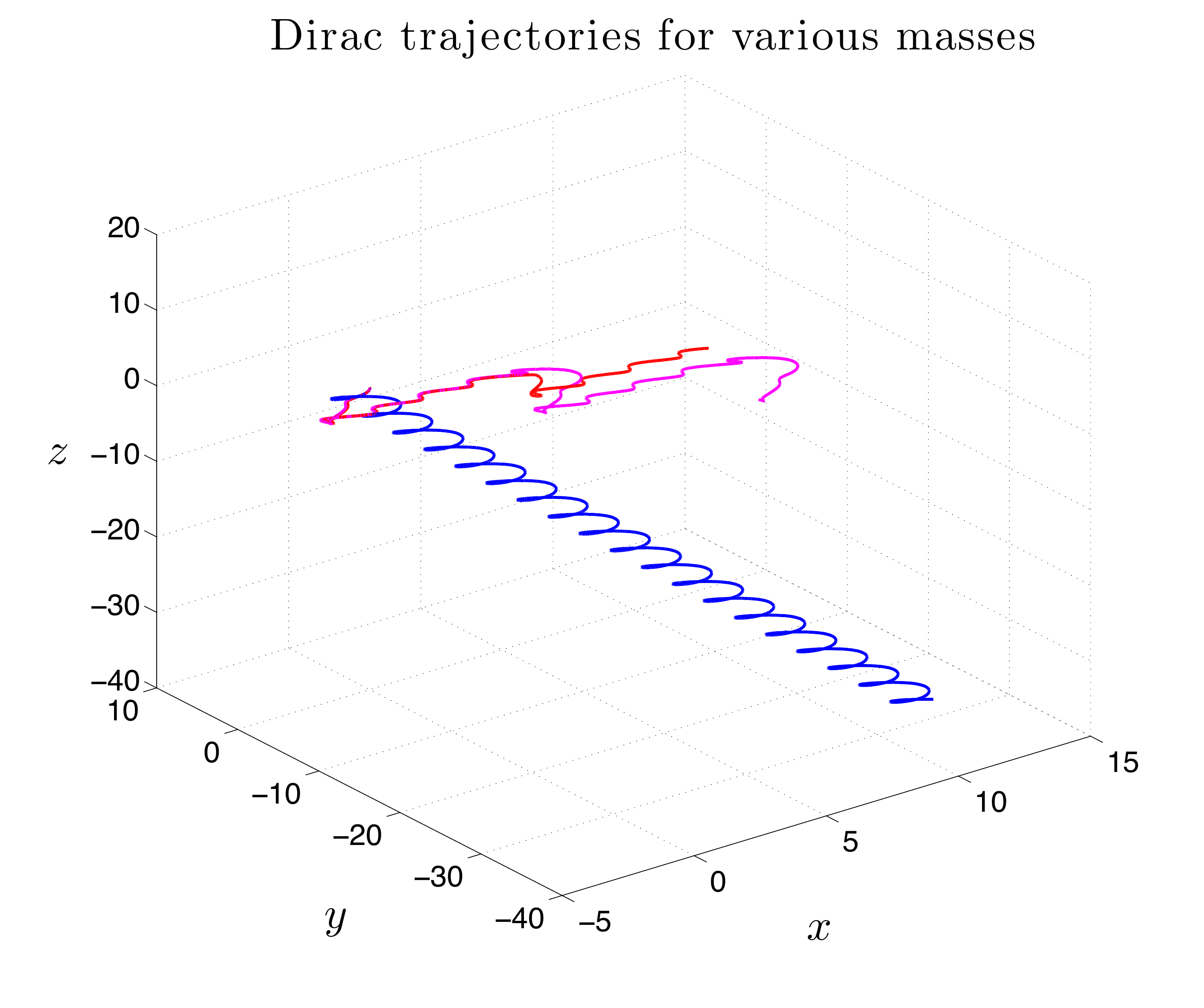}
\caption{\label{fig2}Dirac trajectories for the spinor given in (\ref{dirac_3pw_3d}). 
The 3 trajectories correspond to $3$ different masses ($3$, $6$ and $9$). They all start from the origin and $t\in[0,200]$.}
\end{figure}

\begin{figure}
\includegraphics[width=\textwidth]{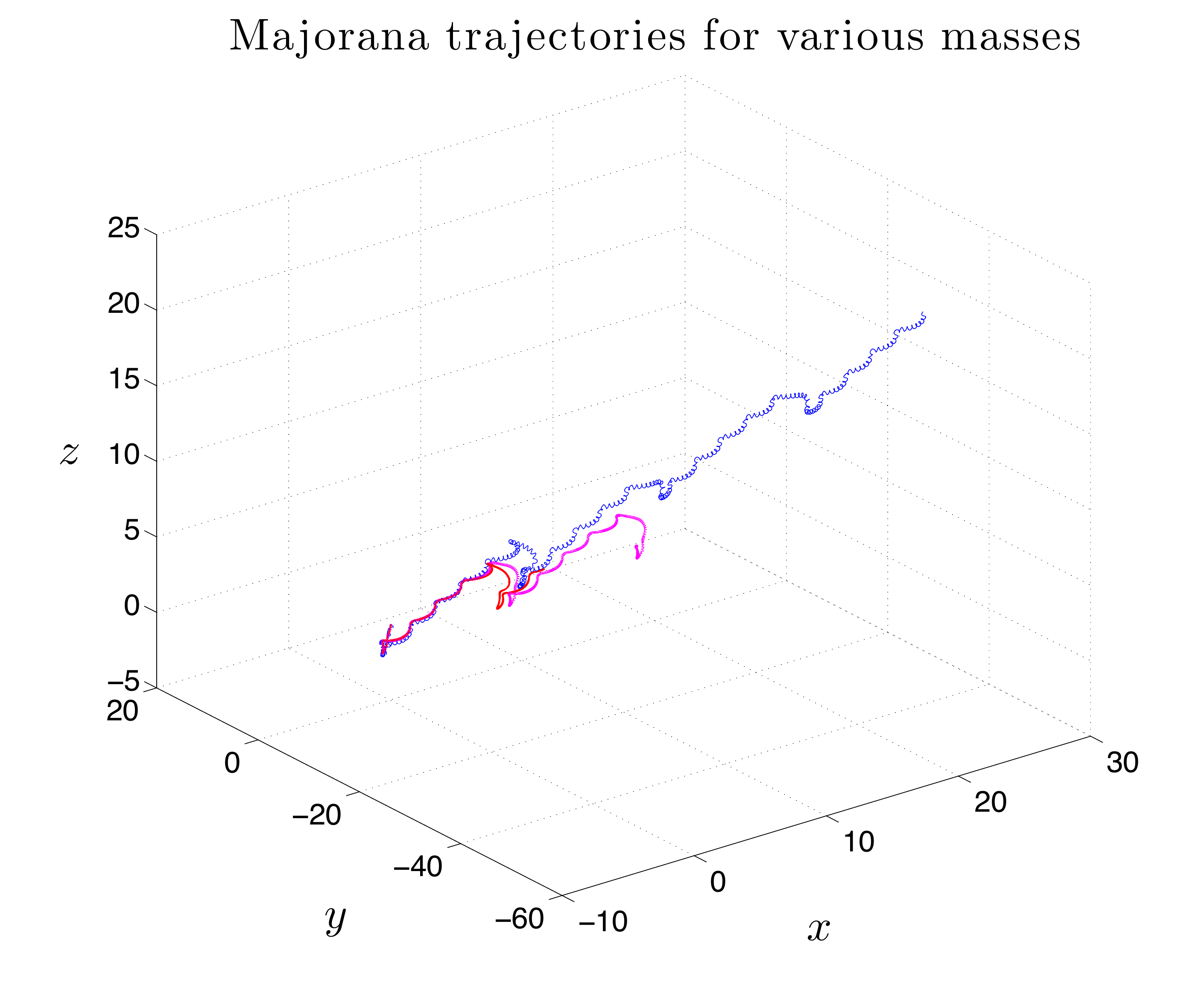}
\caption{\label{fig3} 
Majorana trajectories corresponding to $3$ different masses ($3$, $6$ and $9$). They all start from the origin and $t\in[0,200]$.
The Majorana spinor is obtained from the Dirac spinor given in (\ref{dirac_3pw_3d}), thanks to the construction (\ref{addcc}).}
\end{figure}

\begin{figure}
\includegraphics[width=\textwidth]{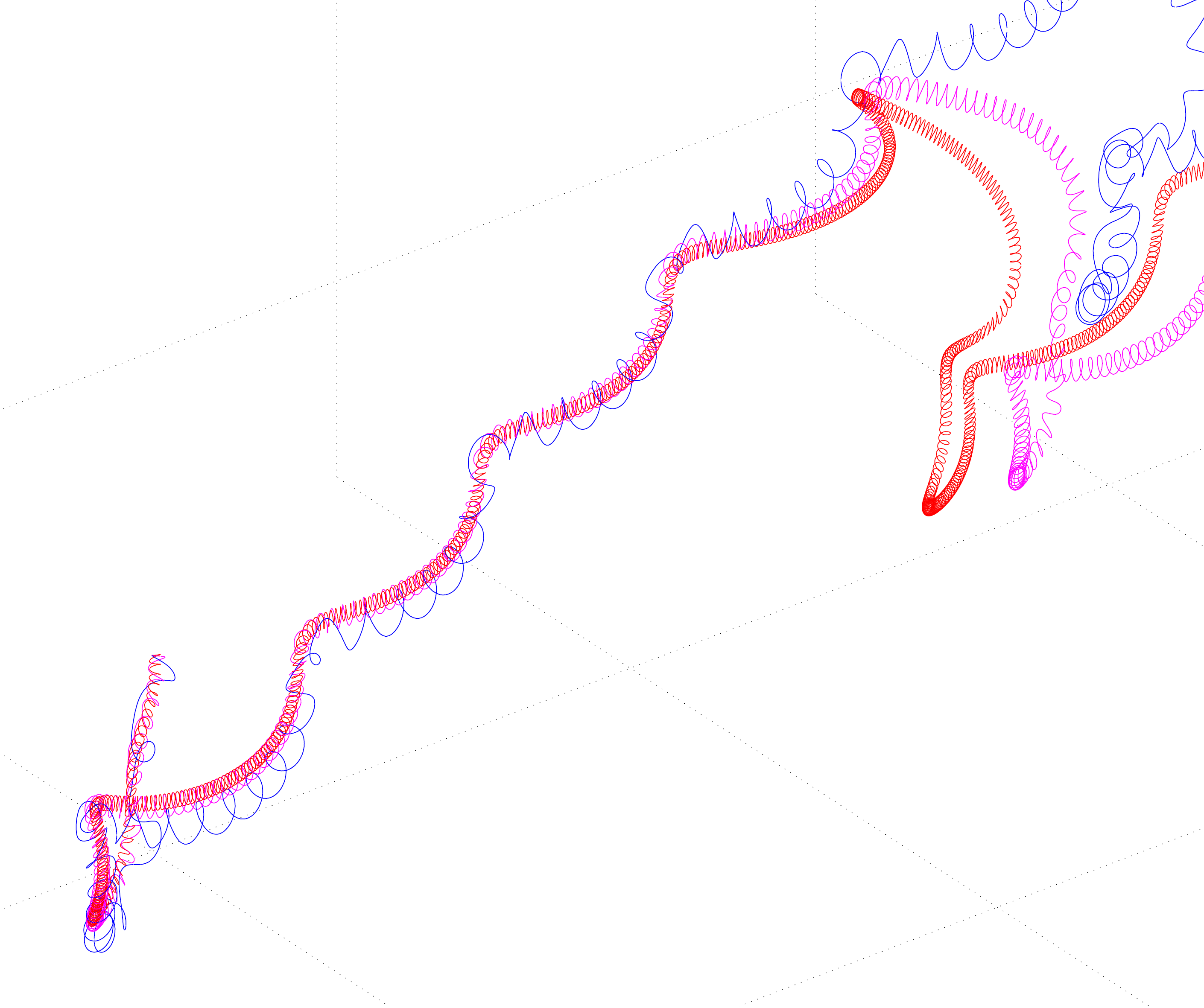}
\caption{\label{fig4}Zoom-in of the trajectories displayed in Figure \ref{fig3}.}
\end{figure}

In order to illustrate the kind of trajectories predicted by the pilot-wave theory in more complex cases, we start from the following Dirac solution
\be\label{dirac_3pw_3d}
\psi_D=\frac{1}{\sqrt{3}}(u_R(\vec{p}_1)e^{-iE_1t+i\vec{p}_1\cdot\vec{x}}+e^{i4}u_R(\vec{p}_2)e^{-iE_2t+i\vec{p}_2\cdot\vec{x}}+e^{i9}u_R(\vec{p}_3)e^{-iE_3t+i\vec{p}_3\cdot\vec{x}})~,\quad\quad
\ee
where
\be 
\vec{p}_1=(1,0,1),~\vec{p}_2=(-1,-2,-1)\textrm{ and }\vec{p}_3=(1,-1,1)~.
\ee 
Then we construct the Majorana spinor $\psi_M=\frac{1}{\sqrt{N}}(\psi_D+i\gamma^2\psi^*_M)$ and solve the guidance equation for a configuration starting at the origin and for different values of the mass. The results are shown in Figure \ref{fig3} (the corresponding results for the Dirac trajectories are shown in Figure \ref{fig2}). Figure \ref{fig4} illustrates the helical nature of the Majorana trajectories.

We make a few observations regarding the Majorana trajectories:
\begin{itemize}
\item We see that the trajectories are helical and this explains how the motion of the configuration can appear subluminal on a coarse-grained level, while being strictly luminal.
\item The larger the mass, the smaller the orbit diameter and the larger the orbit frequency. 
We expect the orbit diameter to be given by the Compton wavelength.
\item The Majorana configuration orbits around a point whose trajectory is subluminal (but which is not the corresponding Dirac trajectory).
\item The helix becomes loose for small masses.
\end{itemize}
\subsection{$2+1$ spacetime}
In a $2+1$ spacetime, the Clifford algebra can be realized with Pauli matrices, for instance $\alpha_1=\sigma_1$, $\alpha_2=\sigma_2$ and $\beta=\sigma_3$. 
Then the Majorana spinor, obtained by adding $\psi_D$ and its charge conjugate $i\sigma_3\sigma_2\psi^*_D$, reads
\be\label{addcc2}
\psi_M=\frac{1}{\sqrt{N}}\begin{pmatrix}\psi_{D,1}+\psi^*_{D,2}\\ \psi_{D,2}+\psi^*_{D,1} \end{pmatrix}~,
\ee
where $\psi_{D,1}$ and $\psi_{D,2}$ are the two components of the Dirac spinor and $N$ is a normalization factor. The motion is luminal again. 

\begin{figure}
\includegraphics[width=\textwidth]{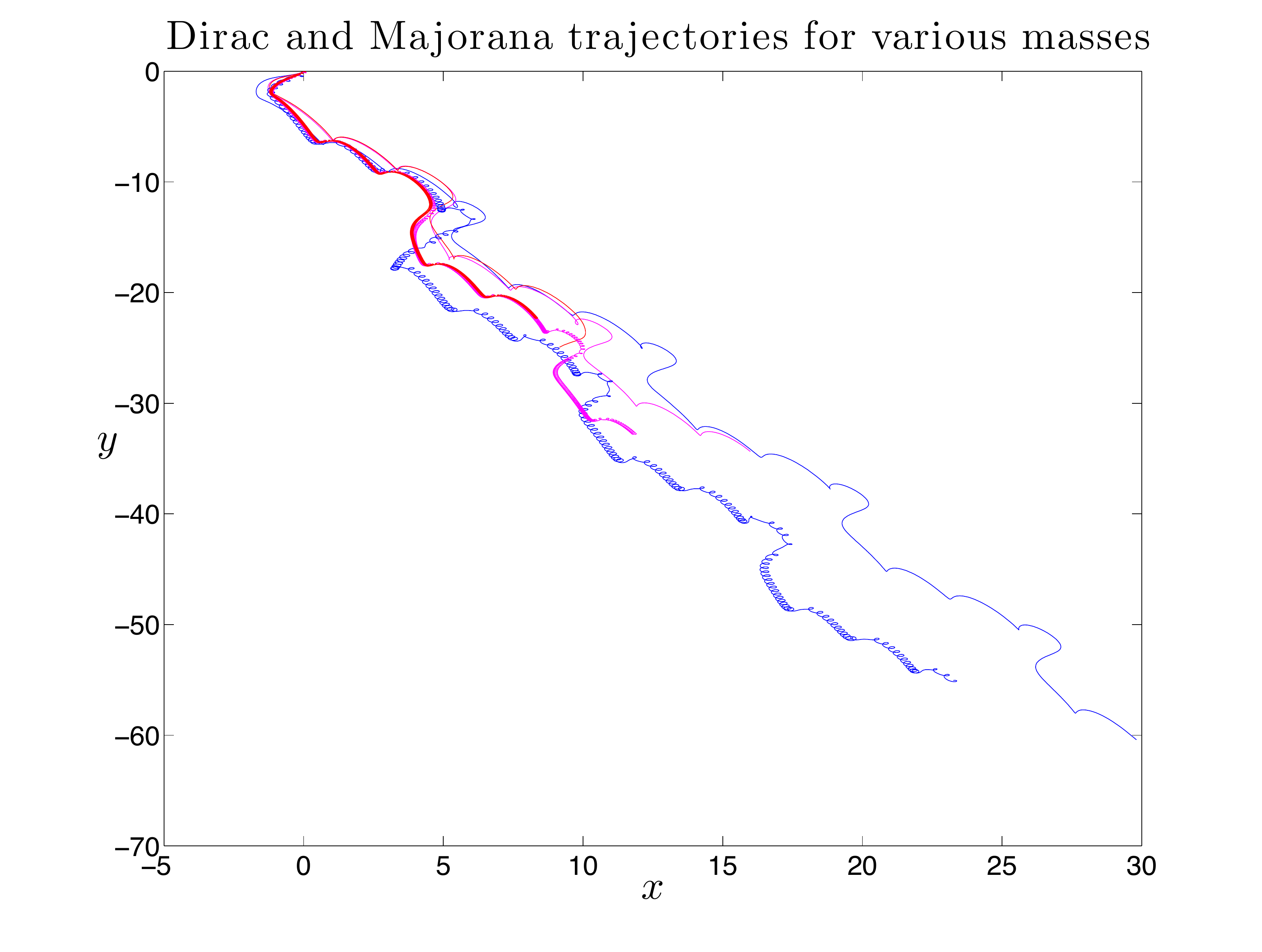}
\caption{\label{fig5}Dirac and Majorana trajectories for various masses in 2D. Masses $\in\{3,6,9\}$. $t\in[0,200]$. The guiding Dirac spinor is given in (\ref{dirac_pw3_2d}). 
The guiding Majorana spinor is obtained from the Dirac spinor thanks to (\ref{addcc2}).}
\end{figure}

As an example, we consider the following Dirac spinor
\be\label{dirac_pw3_2d}
\psi_D=\frac{1}{\sqrt{3}}(u_R(\vec{p}_1)e^{-iE_1t+i\vec{p}_1\cdot\vec{x}}+e^{i4}u_R(\vec{p}_2)e^{-iE_2t+i\vec{p}_2\cdot\vec{x}}+e^{i9}u_R(\vec{p}_3)e^{-iE_3t+i\vec{p}_3\cdot\vec{x}})~,\quad\quad
\ee
where
\be 
\vec{p}_1=(1,0),~\vec{p}_2=(-1,-2)\textrm{ and }\vec{p}_3=(1,-1)~
\ee 
(Basically the same example as before except that the momenta have been truncated). 
We compute the Dirac trajectories and the corresponding Majorana trajectories for three different values of the mass parameter (3,6 and 9). 
As illustrated in Figure \ref{fig5}, the 2D Majorana trajectories share the same properties as the 3D ones.
\subsection{$1+1$ spacetime}
The Majorana motion is not luminal anymore in a $1+1$ spacetime with a two-dimensional representation of the Clifford algebra. 
If it was, non-equilibrium distributions would not relax to quantum equilibrium; indeed the direction of propagation could not be reversed.
\section{Relaxation simulations: general remarks}
Although previous numerical relaxation studies \cite{valentini042,cost10,toruva} support the idea of relaxation to quantum equilibrium, even for Dirac particles \cite{colin2012}, the Majorana trajectory is still very peculiar: it is luminal, it can be helical and if it is helical, the diameter of the helix seems to be equal to the Compton wavelength. 
So it is not altogether impossible that something surprising occurs at the Compton scale when it comes to relaxation to quantum equilibrium. 
For example non-equilibrium may still survive at sub-Compton scales, which would provide a way to partially preserve quantum non-equilibrium. 
If we didn't know about the peculiar nature of the Majorana trajectory, we could argue that the system described by 
$\psi_M={{(\psi_D+i\gamma^2\psi_D^*)}\over\sqrt{N}}$ would have a smaller relaxation time than the system described by $\psi_D$. 
Indeed, increasing the energy spread and the number of modes is assumed to speed up relaxation and $\psi_M$ is just another solution of the Dirac equation, 
but with twice more modes than $\psi_D$ and a considerably larger energy spread than $\psi_D$. 
On the other hand, we could as well argue that $\psi_M$ is more constrained and that goes against chaos (which is essential for a fast relaxation to quantum equilibrium). 
It seems that these questions can only be settled by numerical simulations. 

Thus we want to simulate the temporal evolution of non-equilibrium distributions, for both Dirac and Majorana systems (respectively described by spinors $\psi_D$ and 
$\psi_M=(\psi_D+i\gamma^2\psi_D^*)/\sqrt{N}$), compare the relaxation and draw a conclusion. 
In order to do that, we first need to find a system admitting both Dirac and Majorana solutions. 
Once the system is chosen, we choose a distribution $\rho(t_i,\vec{x})$, different from the Born law distribution, 
and we determine how it evolves thanks to a numerical simulation.

Regarding the choice of the system, the first novelty is that we might be forced to consider  three-dimensional systems instead of standard two-dimensional systems. 
Indeed, considering the simplest case of the Majorana equation in 2D might be unphysical, because spinor nodes can possibly appear in 2D while they are untypical in 3D. 
In the two subsequent sections, we will do both 2D and 3D simulations, and we will see that  nodes, if there are any, do not seem to be an issue.
Now, whether 2D or 3D, we also need a way to confine the fermions. 
A first possible way to achieve this is to use a spherical step potential. However, if we include a spherical potential, 
a solution of the Dirac equation is not a solution anymore under charge conjugation, but we would like to preserve that feature in order to build the Majorana spinors. 
A second option is to consider a sphere and to impose that the normal Dirac flux on the sphere is equal to zero \cite{aldava}, which would confine the fermion 
in the ball inside the sphere (this is also how quarks are confined in the MIT bag model).
However this construction also breaks the invariance under charge conjugation. 
In order to circumvent this problem, we consider a position-dependent mass instead. 
It means that outside of a ball of radius $R$, the mass is given by $M$ while it is given by $m$ inside the ball.  
$M$ will be larger than $m$ but will not tend to infinity.

Regarding the simulation, we use the method developed in \cite{valentini042}, which we summarize in what follows.
In order to compute the density at some final time $t_f$, we use the property that the ratio $\displaystyle\frac{\rho(t,\vec{x})}{|(\psi^\dagger\psi)(t,\vec{x})|}$ is conserved along a trajectory. Then 
$\rho(t_f,\vec{x})$ is given by
\be\label{liou}
\rho(t_f,\vec{x})=(\psi^\dagger\psi)(t_f,\vec{x})\frac{\rho(t_i,BT(\vec{x}))}{(\psi^\dagger\psi)(t_i,BT(\vec{x}))}~,
\ee
where $BT(\vec{x})$ is the initial position which, if evolved according to the guidance equation, would give $\vec{x}$ as the final 
position at $t_f$ ($BT$ stands for backtracking). The relaxation also depends on a coarse-graining (the smaller the coarse-graining length, the longer it would take to see relaxation taking place). In order to perform the coarse-graining, we divide the domain of interest in identical coarse-graining cells. 
The volume (or surface) inside each cell is uniformly sampled by a set of lattice points. For each lattice point $\vec{x}_{lat}$, 
we compute the value  $\rho(t_f,\vec{x}_{lat})$ using (\ref{liou}) and then we average over the set of lattice points contained in the coarse-graining cell, provided 
the backtracking can be trusted. A coarse-grained distribution will be denoted by $\bar{\rho}$.

In the remaining sections, unless specified otherwise, $\psi$ (without index) stands for the Dirac spinor.
\section{Relaxation simulations for a 2D system}
\subsection{Spinor}
We consider a 2D sphere of radius $R$. Inside the sphere, the Dirac spinor behaves as a free spinor of mass $m$.
Outside the sphere, it is a free spinor of mass $M$ instead, with $M>m$. The equation to solve is therefore
\be\label{2dde}
i\partial_t\psi(t,x,y)=-i\alpha_1\partial_x\psi(t,x,y)-i\alpha_2\partial_y\psi(t,x,y)+m(r)\beta\psi(t,x,y)~,
\ee
where $m(r)=m$ if $r=\sqrt{x^2+y^2}\leq R$,  $m(r)=M$ otherwise and where we use the following representation of the matrices: 
$\alpha_1=\sigma_1, \quad\alpha_2=\sigma_2\quad\mathrm{and }\quad\beta=\sigma_3~$. 
In order to obtain the energy eigenstates of this system, we first need to obtain the internal and external energy eigenstates $\psi_{int}$ and $\psi_{ext}$ 
(whose supports are respectively the interior and exterior regions to the sphere). 
Secondly we will take a solution of the form $\psi_{int}+\beta\psi_{ext}$ and 
the boundary condition at $r=R$ will fix $\beta$ and will further restrict the energy to a finite set of discrete eigenvalues. This problem is related 
to the one found in \cite{jafman}.

In order to derive the eigenstates $\psi_{int}$ and $\psi_{ext}$ , we introduce polar coordinates $(r,\theta)$ and we look for positive-energy  eigenstates of the form 
$e^{-iEt}\begin{pmatrix}\psi_1(r,\theta)\\ \psi_2(r,\theta)\end{pmatrix}$. Making these substitutions in (\ref{2dde}), we get the system of equations
\begin{eqnarray}
&(E-\mu)\psi_1&=(-\frac{1}{r}e^{-i\theta}\partial_\theta-i e^{-i\theta}\partial_r)\psi_2\label{sys1}\\
&(E+\mu)\psi_2&=(\frac{1}{r}e^{i\theta}\partial_\theta-i e^{i\theta}\partial_r)\psi_1\label{sys2}~,
\end{eqnarray}
where $\mu$ either stands for $m$ or $M$ depending on whether we look for internal or external solutions.
The solutions are therefore of the form $\psi_{n,1}=\tilde{\psi}_{n,1}(r) e^{i n\phi}$ and $\psi_{n,2}=\tilde{\psi}_{n,2}(r) e^{i(n+1)\phi}$, with n a positive integer, or of the form 
$\psi_{p,1}=\tilde{\psi}_{p,1}(r)e^{-i p\phi}$ and $\psi_{p,2}=\tilde{\psi}_{p,2}e^{i(p-1)\phi}$ with $p$ a strictly positive integer. 
The system of equations becomes
\begin{eqnarray}
(E^2-\mu^2)\tilde{\psi}_{n,1}=(\frac{n^2}{r^2}-\frac{1}{r}\partial_r-\partial^2_r)\tilde{\psi}_{n,1}\label{sysn1}\\
\tilde{\psi}_{n,2}=\frac{i}{E+\mu}(\frac{n}{r}-\partial_r)\tilde{\psi}_{n,1}\label{sysn2}
\end{eqnarray}
and 
\begin{eqnarray}
(E^2-\mu^2)\tilde{\psi}_{p,1}=(\frac{p^2}{r^2}-\frac{1}{r}\partial_r-\partial^2_r)\tilde{\psi}_{p,1}\label{sysp1}\\
\tilde{\psi}_{p,2}=\frac{-i}{E+\mu}(\frac{p}{r}+\partial_r)\tilde{\psi}_{p,1}\label{sysp2}~.
\end{eqnarray}
The solutions of (\ref{sysn1}) and (\ref{sysp1}) are Bessel functions but which exact ones depends on the sign of $(E^2-\mu^2)$. If $(E^2-\mu^2)$ is positive, 
the solutions are of the form $J_n(\sqrt{E^2-\mu^2}r)$ or $Y_n(\sqrt{E^2-\mu^2}r)$. If $(E^2-\mu^2)$ is negative, 
the solutions are $I_n(\sqrt{\mu^2-E^2}r)$ or $K_n(\sqrt{\mu^2-E^2}r)$. In order to further restrict the possible solutions, 
we shall assume that $E\in]m,M[$. If we do that, internal solutions can only be of the type $J$ whereas outside solutions can only be of the type $K$, the reason 
being that the $Y's$ are not regular at the origin while the $I's$ blow up at infinity. $\tilde{\psi}_{n,2}$ and $\tilde{\psi}_{p,2}$ can be obtained from $\tilde{\psi}_{n,1}$ and 
$\tilde{\psi}_{p,1}$ thanks to (\ref{sysn2}) and (\ref{sysp2}) and they can be further simplified by using the recurrence relations satisfied 
by the Bessel functions $J$ and $K$. Overall the solutions read
\be
\psi^{int}_{n}=\begin{pmatrix} J_{n}(k_{int} r)e^{i n \theta}\\ i\displaystyle\frac{k_{int}}{E+m}J_{n+1}(k_{int} r)e^{i(n+1)\theta}\end{pmatrix}~
\psi^{ext}_{n}=\begin{pmatrix} K_{n}(k_{ext} r)e^{i n \theta}\\ i\displaystyle\frac{k_{ext}}{E+M}K_{n+1}(k_{ext} r)e^{i(n+1)\theta}\end{pmatrix}
\ee
and
\be
\psi^{int}_{p}=\begin{pmatrix} J_{p}(k_{int} r)e^{-i p \theta}\\ -i\displaystyle\frac{k_{int}}{E+m}J_{p-1}(k_{int} r)e^{-i(p-1)\theta}\end{pmatrix}~
\psi^{ext}_{p}=\begin{pmatrix} K_{p}(k_{ext} r)e^{-i p \theta}\\ i\displaystyle\frac{k_{ext}}{E+M}K_{p-1}(k_{ext} r)e^{-i(p-1)\theta}\end{pmatrix}
\ee
with $k_{int}=\sqrt{E^2-m^2}$ and $k_{ext}=\sqrt{M^2-E^2}$.

Now we take a solution of the form $\psi^{int}+\beta\psi^{ext}$ and we impose that
$\frac{\psi^{int}_1}{\psi^{int}_{2}}\bigg|_{r=R}=\frac{\beta}{\beta}\frac{\psi^{ext}_1}{\psi^{ext}_{2}}\bigg|_{r=R}$~,
which give the equations
\be
(E+m)\frac{J_n(k_{int}R)}{k_{int} J_{n+1}(k_{int}R)}=(E+M)\frac{K_n(k_{ext}R)}{k_{ext} K_{n+1}(k_{ext}R)}
\ee
and
\be
-(E+m)\frac{J_p(k_{int}R)}{k_{int} J_{p-1}(k_{int}R)}=(E+M)\frac{K_p(k_{ext}R)}{k_{ext} K_{p-1}(k_{ext}R)}~.
\ee
The solutions $E$, which can only be found numerically, correspond to the discrete set of energy eigenvalues. Once $E$ is known, the corresponding 
$\beta$ is fixed thanks to $J_n(k_{int} R)=\beta K_n(k_{ext}R)$ or $J_p(k_{int} R)=\beta K_p(k_{ext}R)$.
Finally each solution is normalized, the useful relation \cite{jeffrey_dai} being $\int x J^2_{n}(ax)dx=\frac{x^2}{2}\left(J^2_n(ax)-J_{n-1}(a x) J_{n+1} (a x)\right)~$.
\subsection{Simulations}
\begin{figure}
\includegraphics[width=\textwidth]{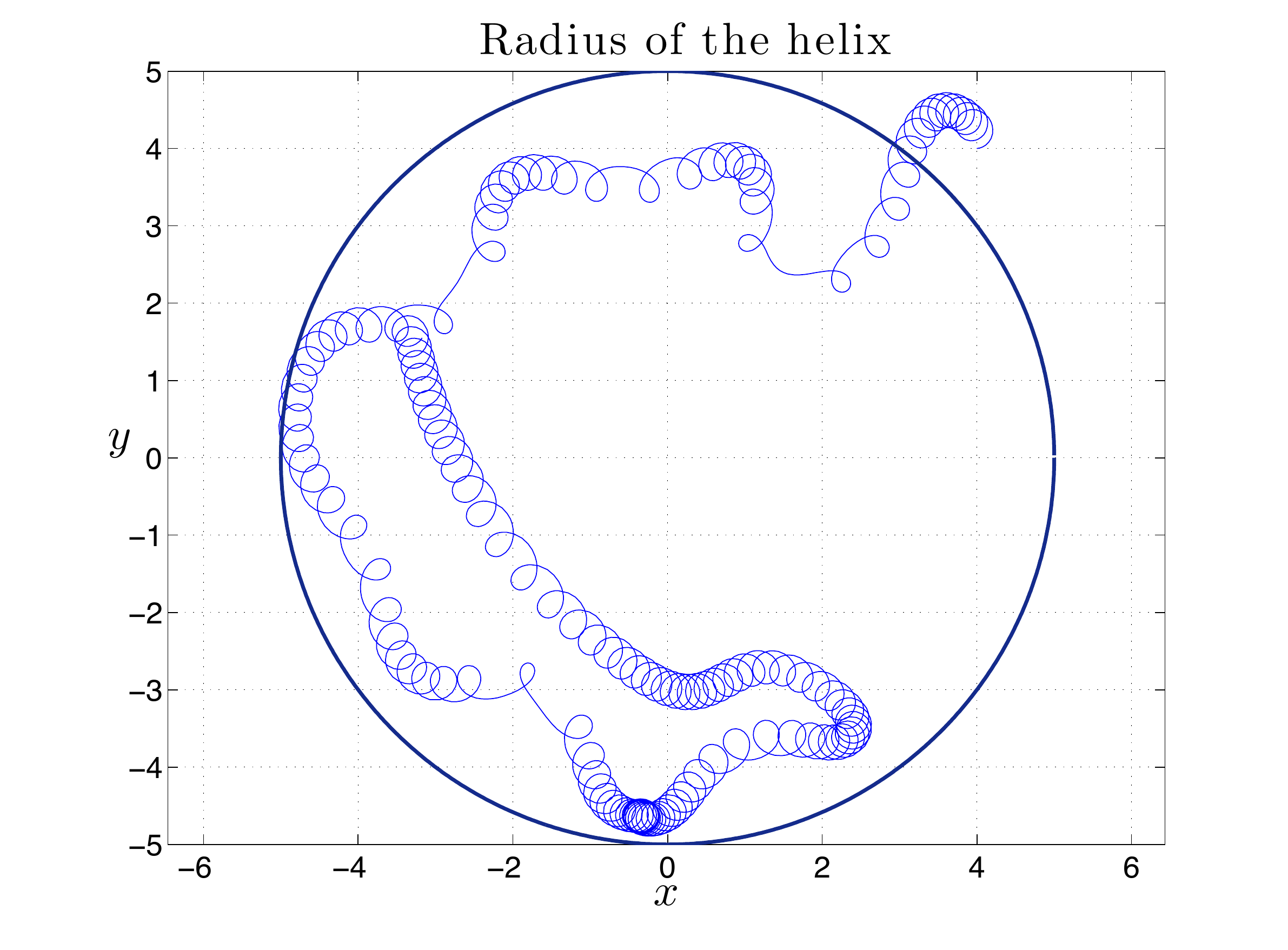}
\caption{\label{fig6}Majorana trajectory originating from $(4,4)$ at $t=0$.
The Majorana spinor is obtained from the Dirac spinor defined in Table \ref{spinor2d} thanks to the construction defined at (\ref{addcc}).}
\end{figure}

We have chosen $R=5$, $m=2$ and $M=2.5$. Thus we expect, if we still have a helix, that its inside diameter will be approximately equal to 0.5. 
We have chosen $M$ relatively close to $m$ in order for $E\in]m,M[$ to be close to $m$ as well (that should give a 
strong helix hopefully). But this will restrict the number of available modes as well because the domain of the energy is restricted to $]2,2.5[$. We were 
able to find $6$ modes satisfying these conditions. These six modes are superposed with equal weight coefficients in order to build the full spinor; details for the superposition are given in Table \ref{spinor2d}.
\begin{table}
\begin{center}
\begin{tabular}{| c| r| r| c | c|}
\hline
$\textrm{mode}$ &  $qn$ & $j$ &  $\textrm{energy}$ & $\textrm{phase}$\\
\hline
$1$ & $0~(n=0)$ &  $1/2$  & $2.0431085410058$ & $e^{i5.11905989575681}$\\
\hline
$2$ & $1~(n=1)$ &  $3/2$  & $2.10705432759443$ & $e^{i5.69125859039527}$\\
\hline
$3$ & $2~(n=2)$ &  $5/2$  & $2.18732348257316$ & $e^{i0.79788169834087}$\\
\hline
$4$ & $-1~(p=1)$ &  $-1/2$  & $2.10790439723629$ & $e^{i5.73890975922526}$\\
\hline
$5$ & $-2~(p=2)$ &  $-3/2$  & $2.19010176106309$ & $e^{i3.97323032474265}$\\
\hline
$6$ & $-3~(p=3)$ &  $-5/2$  & $2.2864073103276$ & $e^{i0.61286443954863}$\\
\hline
\end{tabular}
\caption{\label{spinor2d}Definition of the 6 modes whose sum, with equal weight coefficients, gives the full 2D Dirac spinor.}
\end{center}
\end{table}
Once the spinor is implemented, we can check that we have indeed a helix and that its diameter is approximately equal to $0.5$; this is illustrated in Figure \ref{fig6}.

\begin{figure}
\includegraphics[width=\textwidth]{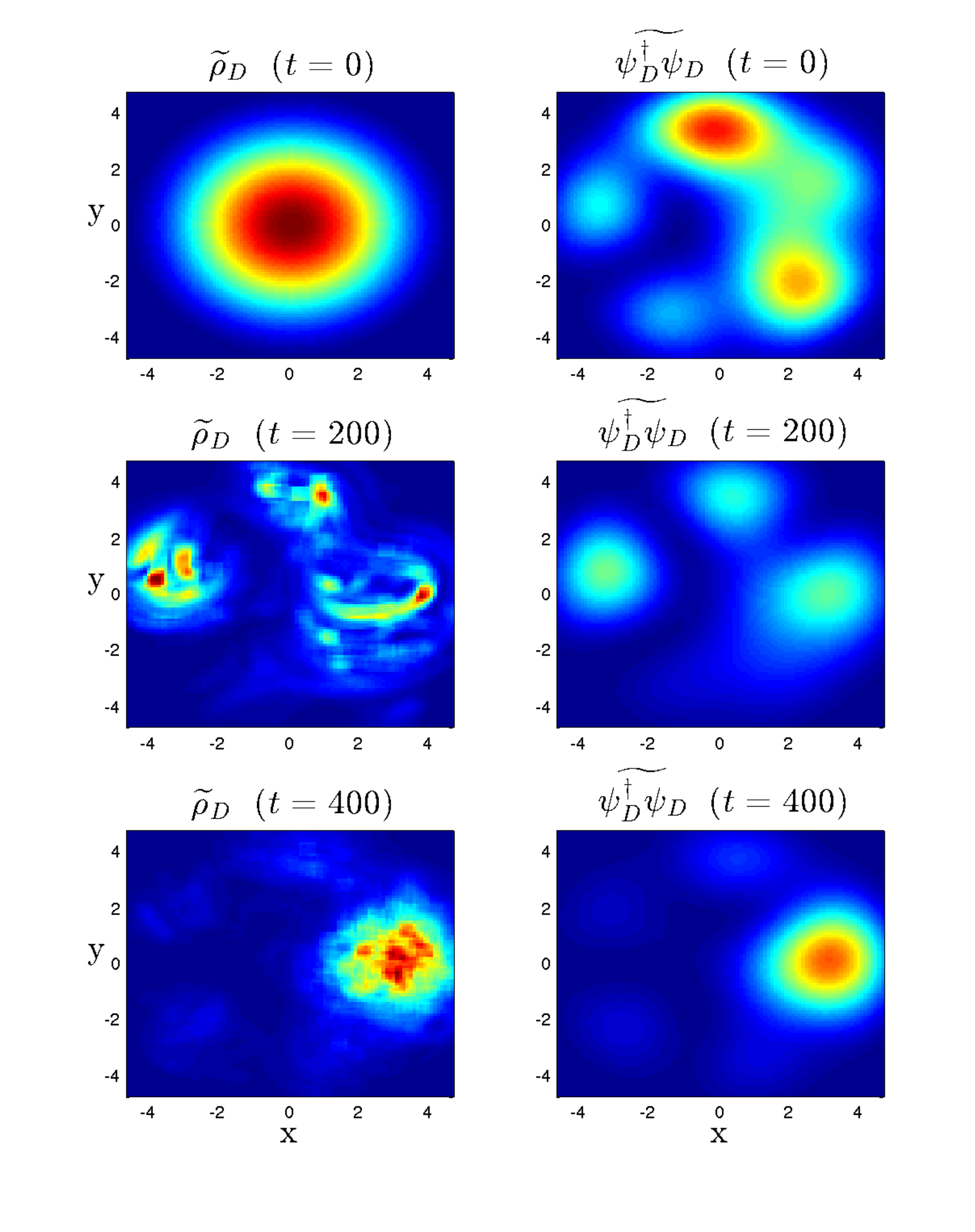}
\caption{\label{fig7}Evolution of $\widetilde\rho_{D}$ and $\widetilde{\psi^\dagger_D\psi_D}$. The initial non-equilibrium distribution is defined 
by $\rho_D(t=0,r,\theta)=\rho_1(r,\theta)$, where $\rho_1$ is defined in (\ref{rho02d}). 
The Dirac spinor is defined in Table \ref{spinor2d}.}
\end{figure}

\begin{figure}
\includegraphics[width=\textwidth]{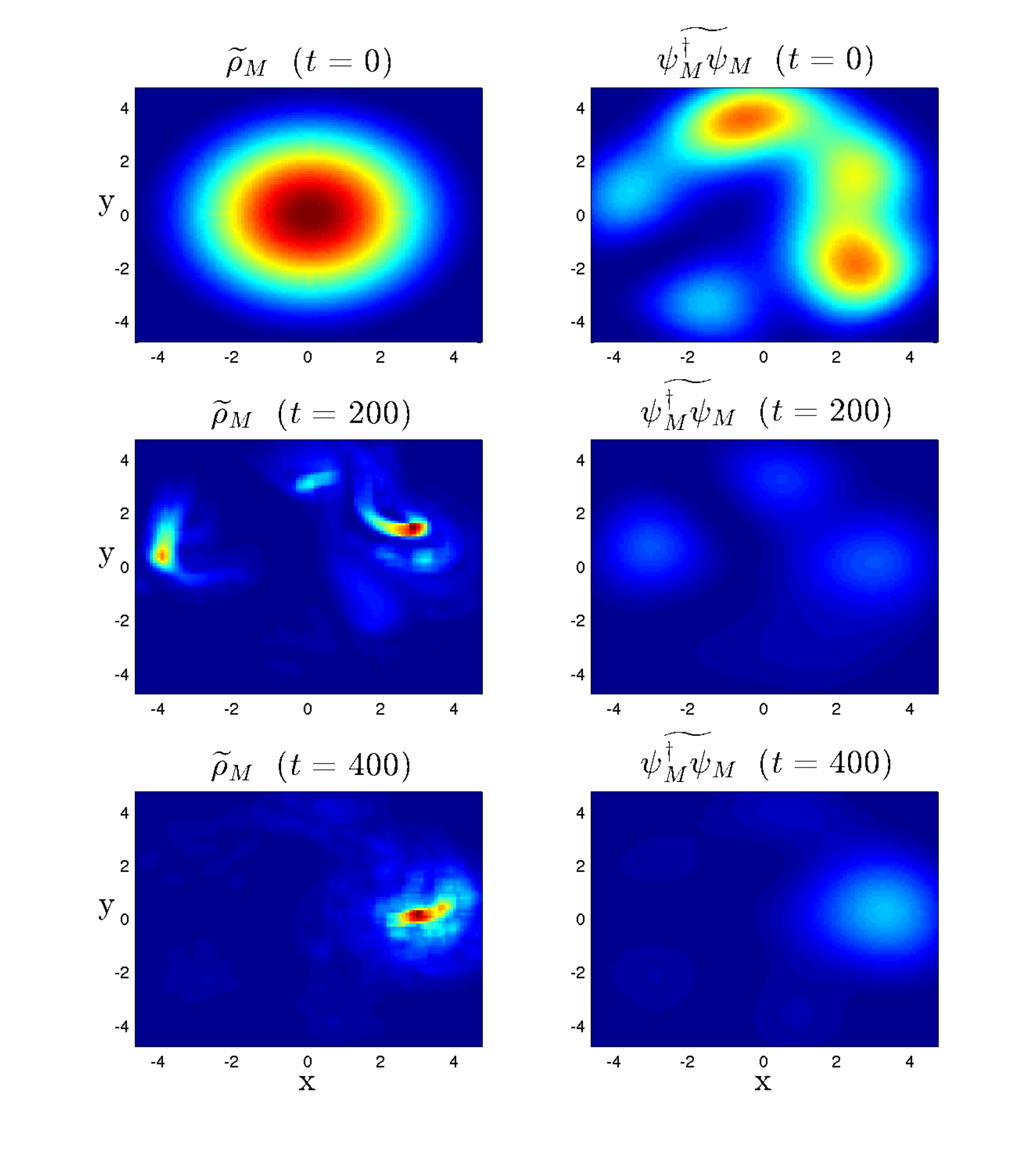}
\caption{\label{fig8}Evolution of $\widetilde\rho_{M}$ and $\widetilde{\psi^\dagger_M\psi_M}$. The initial non-equilibrium distribution is defined 
by $\rho_M(t=0,r,\theta)=\rho_1(r,\theta)$, where $\rho_1$ is defined in (\ref{rho02d}). 
The Majorana spinor is obtained from the Dirac spinor defined in Table \ref{spinor2d} thanks to the construction defined at (\ref{addcc}).}
\end{figure}

The initial non-equilibrium is defined as follows
\be\label{rho02d}
\rho_1(r,\theta)=2\pi\frac{\cos^2(\frac{\pi r}{2 R})}{R^2(\pi^2-4)}~\textrm{if }r\leq R\textrm{~and 0 otherwise}.
\ee
For the relaxation simulations, we have looked at the region $[-5,5]\times[-5,5]$. We have used a grid of $400\times400$ points 
covering that domain (the points belonging to that grid have positions $(-5+k\frac{10}{400}-\frac{10}{800},-5+l\frac{10}{400}-\frac{10}{800})$ with $k,l\in\{1,2,\ldots,400\}$ ). The domain is divided in $20\times20$ non-overlapping coarse-graining cells (each coarse-graining cell contains $400$ points). For the plots, we use $96\times96$ 
overlapping coarse-graining cells (each of these cell also contains 400 points); a smooth coarse-graining of $\rho$, using overlapping coarse-graining cells, is 
denoted by $\widetilde{\rho}$. The results are plotted in Figures \ref{fig7} and \ref{fig8}.
\subsection{Discussion}
The relaxation of the non-equilibrium distribution in the Majorana case (Figure \ref{fig8}) is clearly retarded with respect to the Dirac case (Figure \ref{fig7}). 
This retardation cannot be attributed to the difference in $\psi^\dagger_D\psi_D$ and $\psi^\dagger_M\psi_M$ at the initial time because they are approximately equal.
In order to pinpoint 
the origin of the delay, we will compute, in both cases, 5 trajectories whose initial positions are close or equal to $(0,3)$ (this position is the centre of a region where 
$\psi^\dagger_D\psi_D$, $\psi^\dagger_M\psi_M$ and $\rho$ as well have ``good'' support initially).

\begin{figure}
\includegraphics[width=\textwidth]{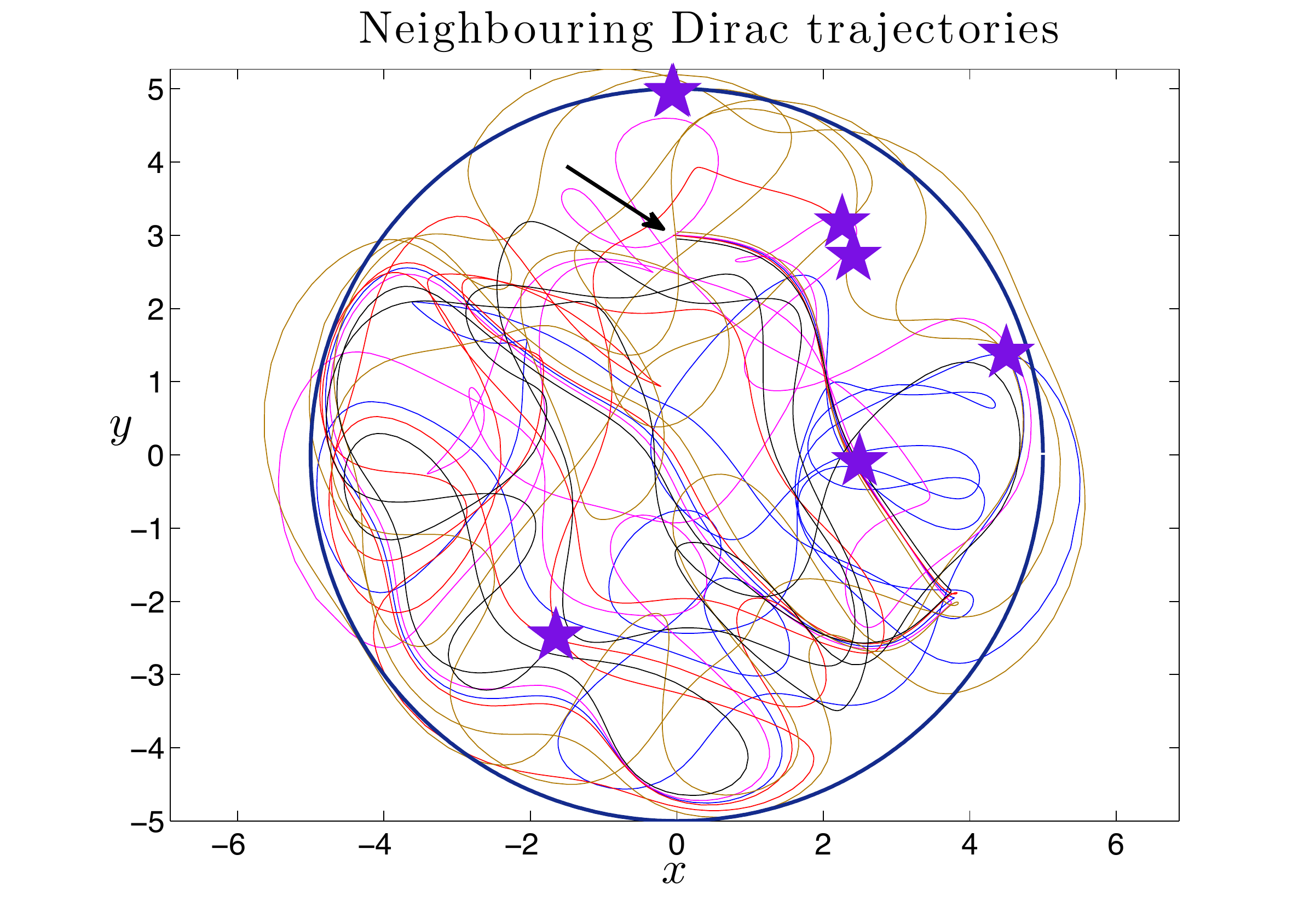}
\caption{\label{fig9}Five Dirac trajectories originating, at $t=0$, from $(0,3)$, $(0,3.05)$, $(0,2.95)$, $(-0.05,3)$ and $(0.05,3)$.
The Dirac spinor is defined in Table \ref{spinor2d}. $t\in[0,200]$. 
The purple stars indicate the final positions whereas the black arrow indicates $(0,3)$.}
\end{figure}

\begin{figure}
\includegraphics[width=\textwidth]{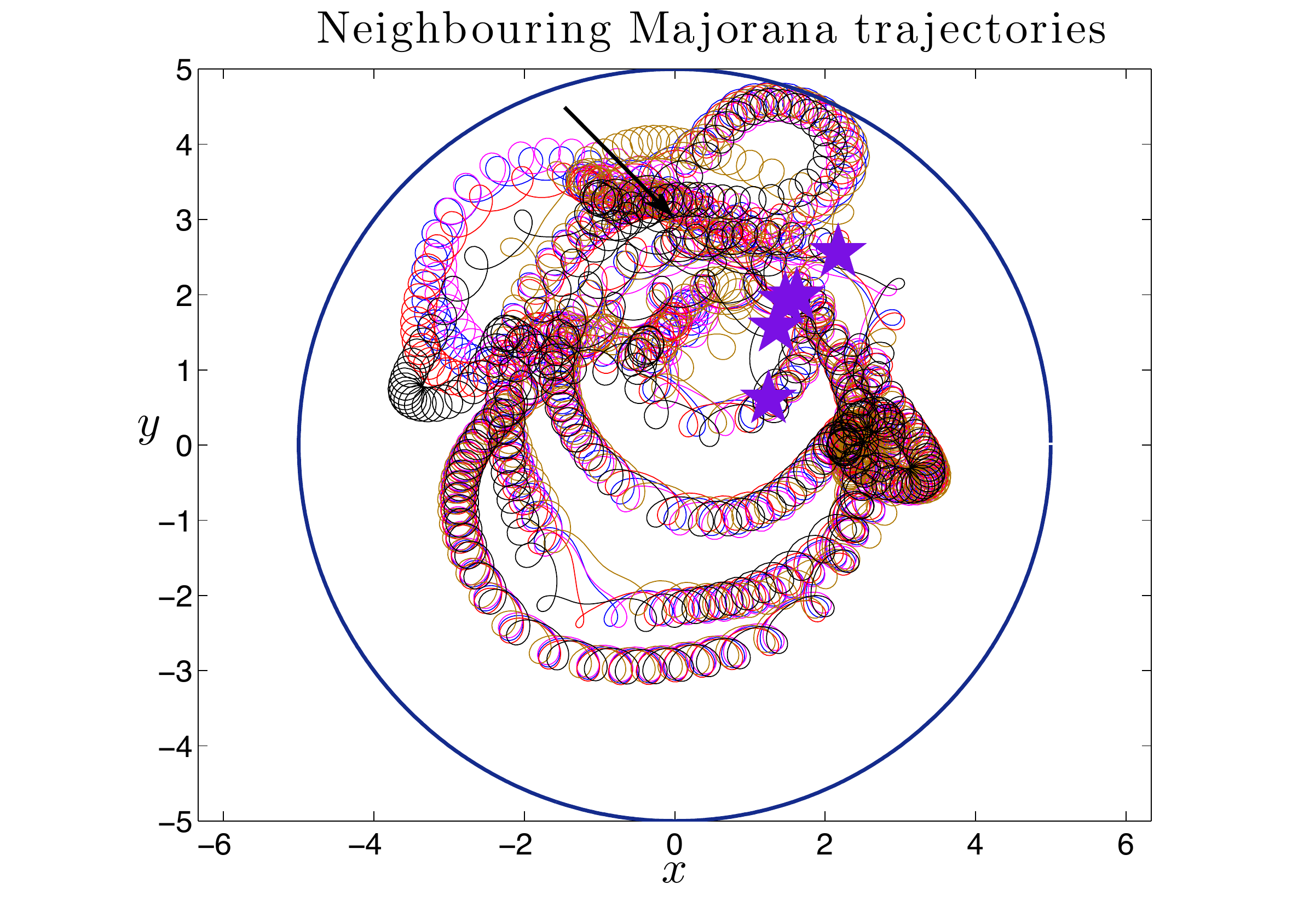}
\caption{\label{fig10}
Five Majorana trajectories originating from $(0,3)$, $(0,3.05)$, $(0,2.95)$, $(-0.05,3)$ and $(0.05,3)$ at $t=0$.
The Majorana spinor is obtained from the Dirac spinor defined in Table \ref{spinor2d} by adding the Dirac spinor and its charge conjugate. $t\in[0,200]$. 
The purple stars indicate the final positions whereas the black arrow indicates $(0,3)$.}
\end{figure}

As it is illustrated in Figures \ref{fig9} and \ref{fig10}, Majorana trajectories originating from neighbouring points, in contrast to Dirac trajectories, tend to follow the same helix 
and to remain close together. This is bad for relaxation because they behave, roughly speaking, as a single trajectory of width equal to the Compton wavelength. 
And it is known that relaxation to quantum equilibrium doesn't hold at the level of a single trajectory (this a consequence of (\ref{liou})) and therefore 
a fast divergence of neighbouring trajectories is necessary for a fast relaxation to quantum equilibrium. 

\section{Relaxation simulations for a 3D system}
\subsection{Spinor}
We want to find the solutions of 
\be
i\partial_t\psi(t,\vec{x})=-i\vec{\alpha}\cdot\vec{\nabla}\psi(t,\vec{x})+m(r)\beta\psi(t,\vec{x})~,
\ee
where $m(r)=M$ if $r\geq R$ and $m(r)=m$ otherwise. 
This problem is a special case of a larger class of problems treated in the literature (single Dirac particle in a central potential, see \cite{dirac,sakurai,rafelski}). 
Actually, in our case, we only have a spherical boundary condition at $r=R$ but no potential. In the next paragraph, we indicate the basic steps for the derivation 
of the energy eigenstates.
 
The solutions will be characterized by four quantum numbers corresponding to the operators $H$, $K$, $J$ and $J_3$, 
where $K$ is the Dirac operator 
\be
K=\begin{pmatrix}\vec{\sigma}\cdot\vec{L}+\hbar & 0\\0 & -\vec{\sigma}\cdot\vec{L}-\hbar\end{pmatrix}~,
\ee
$\vec{L}$ the angular momentum and $\vec{J}$ the total angular momentum. The eigenvalues of $H$, $K$, $J^2$ and $J_3$ are 
respectively denoted by $E$, $-\kappa\hbar$, $j(j+1)\hbar^2$ and $j_3\hbar$. $\kappa$ is a non-zero integer and it must be related to $j$ by the following relation
\be
\kappa=\pm(j+\frac{1}{2})~.
\ee
The spinor can be written as
\be
\psi=\begin{pmatrix}\psi_A\\\psi_B\end{pmatrix}=\begin{pmatrix} g(r)\mathcal{Y}^{j_3}_{j l_A} \\ if(r)\mathcal{Y}^{j_3}_{j l_B}\end{pmatrix}
\ee
where $l_A$ and $l_B$ are fixed by $j$ and $\kappa$ (if $\kappa=j+1/2$, then $l_A=j+1/2$, $l_B=j=1/2$ and vice-versa if $\kappa=-j-1/2$). 
If $j=l+1/2$ then 
\be
\mathcal{Y}^{j_3}_{jl}=\sqrt{\frac{l+j_3+1/2}{2l+1}} Y^{j_3-\frac{1}{2}}_{l}\begin{pmatrix}1\\0\end{pmatrix}+
\sqrt{\frac{l-j_3+1/2}{2l+1}} Y^{j_3+\frac{1}{2}}_{l}\begin{pmatrix}0\\1\end{pmatrix}~,
\ee
otherwise ($j=l-1/2$)
\be
\mathcal{Y}^{j_3}_{jl}=-\sqrt{\frac{l-j_3+1/2}{2l+1}} Y^{j_3-\frac{1}{2}}_{l}\begin{pmatrix}1\\0\end{pmatrix}+\sqrt{\frac{l-j_3+1/2}{2l+1}} Y^{j_3+\frac{1}{2}}_{l}
\begin{pmatrix}0\\1\end{pmatrix}~
\ee
where the $Y$ are the usual spherical harmonics with the Condon-Shortley phase convention. 
For an energy eigenstate with eigenvalue $E$, $f(r)$ and $g(r)$ are solutions of the system of equations
\begin{eqnarray}
f=\frac{1}{E+m(r)}(\frac{dg}{dr}+\frac{1+\kappa}{r}g)\label{fg}\\
-\frac{d^2g}{dr^2}-\frac{2}{r}\frac{dg}{dr}+\frac{\kappa(\kappa+1)}{r^2}g=(E^2-m^2(r))g~.\label{bessel}
\end{eqnarray}
If $E^2-m^2(r)>0$, $g$ will be a spherical Bessel function, otherwise ($E^2-m^2(r)<0$) $g$ will be a modified spherical Bessel function. 
We are going to superpose energy eigenstates with eigenvalues $E$ such that
\be
E^2-m^2>0~\quad\mathrm{and}~\quad M^2-E^2>0~.
\ee
Under these conditions, if we look at (\ref{bessel}), we see that the internal solution is given by 
\be
g_{int}(r)=j_{l(\kappa)}(p_{int}r)
\ee
where $p_{int}=\sqrt{E^2-m^2}$, $l(\kappa)=\kappa$ if $\kappa> 0$, $l(\kappa)=-\kappa-1$  if $\kappa<0$, and $j_{l(\kappa)}$ is a spherical Bessel function of the first kind (the only one regular at the origin). 
On the other hand, the external solution is given by 
\be
g_{ext}(r)=k_{l(\kappa)}(p_{ext}r)
\ee
where $p_{ext}=\sqrt{M^2-E^2}$ and $k_{l(\kappa)}$is a modified spherical Bessel function of the second kind. If we put these solutions in (\ref{fg}), we find out, using 
the recurrence relations for Bessel functions, that 
\be
f_{int}(r)=\frac{\kappa}{|\kappa|}\frac{p_{int}}{E+m}j_{lm(\kappa)}(p_{int}r)\quad\quad f_{ext}(r)=-\frac{p_{ext}}{E+M}k_{lm(\kappa)}(p_{ext}r)~,
\ee
where $lm(\kappa)=\kappa-1$ if $\kappa>0$ and $lm(\kappa)=-\kappa$ if $\kappa<0$.
Overall we have that 
\be
\psi_{int}=Ae^{-iEt}\begin{pmatrix}j_{l(\kappa)}(p_{int}r)\mathcal{Y}^{j_3}_{j l_A}(\theta,\phi)\\i\frac{\kappa}{|\kappa|}\frac{p_{int}}{E+m}j_{lm(\kappa)}(p_{int}r)\mathcal{Y}^{j_3}_{j l_B}(\theta,\phi)\end{pmatrix}
\ee
and 
\be
\psi_{ext}=Be^{-iEt}\begin{pmatrix}j_{l(\kappa)}(p_{ext}r)\mathcal{Y}^{j_3}_{j l_A}(\theta,\phi)\\-i\frac{p_{ext}}{E+M}j_{lm(\kappa)}(p_{ext}r)\mathcal{Y}^{j_3}_{j l_B}(\theta,\phi)\end{pmatrix}
\ee
where $A$ and $B$ are two constants, which can be fixed by the normalization of the spinor and by matching the upper components of the internal and external spinors 
at the boundary $r=R$. But first one needs to find the energy eigenvalues. 
This can be done by matching the ratios $\frac{\psi_A}{\psi_B}$ for the internal and external solutions at the boundary.
The good thing about the types of Bessel functions involved in the above expressions is that 
their analytic expressions are known, for example:
\be
j_1(x)=\frac{\sin{x}}{x^2}-\frac{\cos{x}}{x}\textrm{ and }k_1(x)=\frac{e^{-x}}{x^2}(1+x)
\ee
\subsection{Simulations}
We have obtained the analytical expressions for the energy eigenstates. 
The next step is to choose some numerical values for the masses, for the radius $R$, choose a few modes 
with quantum numbers $\kappa$, $j$ and $j_3$ and determine the corresponding energy eigenvalues. This last part has to be done numerically. 
As in the 2D case, we have taken $R=5$, but the two masses are now given by $m=1$ and $M=1.5$. 
The coarse-graining length is equal to 1 and the parameters used for the superposition can be found in Table \ref{spinor3d}.
\begin{table}
\begin{center}
\begin{tabular}{| c| c| c| c | c| c|}
\hline
$\textrm{mode}$ &  $\kappa$ & $j$ &  $j_3$ & $\textrm{energy}$ & $\textrm{phase}$\\
\hline
$1$ & $1$ &  $1/2$  & $1/2$ & $1.24382856361355$ & $e^{i5.11905989575681}$\\
\hline
$2$ & $1$ &  $1/2$  & $-1/2$ & $1.24382856361355$ & $e^{i5.69125859039527}$\\
\hline
$3$ & $-1$ &  $1/2$  & $1/2$ & $1.12290254936835$ & $e^{i0.79788169834087}$\\
\hline
$4$ & $-1$ &  $1/2$  & $-1/2$ & $1.42209141193299$ & $e^{i5.73890975922526}$\\
\hline
$5$ & $2$ &  $3/2$  & $-1/2$ & $1.37899955258739$ & $e^{i3.97323032474265}$\\
\hline
$6$ & $2$ &  $3/2$  & $3/2$  & $1.37899955258739$ & $e^{i0.61286443954863}$\\
\hline
$7$ & $-2$ &  $3/2$  & $-3/2$ & $1.23582476498988$ & $e^{i1.74985591686112}$\\
\hline
$8$ & $-2$ &  $3/2$  & $1/2$ & $1.23582476498988$ & $e^{i3.43615792623681}$\\
\hline
\end{tabular}
\caption{\label{spinor3d}Definition of the 8 modes whose sum, with equal weight coefficients, gives the full 3D Dirac spinor.}
\end{center}
\end{table}
Here is an example of trajectory for this system, in order to illustrate that the motion is indeed helical with 
a helix diameter approximately equal to the Compton wavelength.
\begin{figure}
\centering
\includegraphics[width=\textwidth]{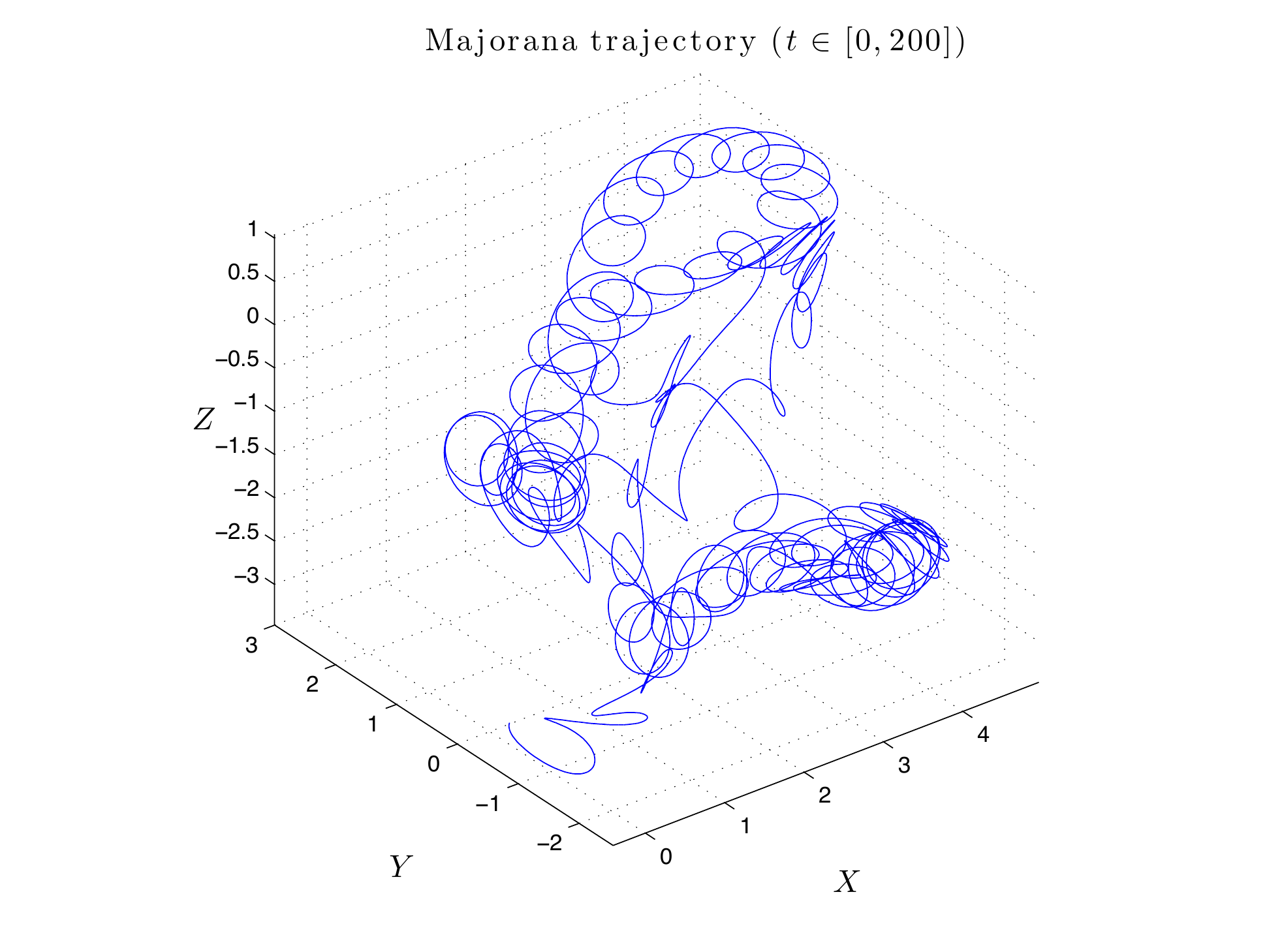}
\caption{\label{fig11}Majorana trajectory starting from the position $(1,-0.5,0.7)$ illustrating the helical nature of the trajectory. 
The spinor guiding the beable is defined in Table \ref{spinor3d}. $t\in[0,200]$.}
\end{figure}
The initial non-equilibrium is defined as follows
\be\label{rho03d}
\rho_2(r,\theta,\phi)=\cos{(0.5\frac{\pi r}{R})}\frac{\pi^2}{8 R^3(\pi^2-8)}~\textrm{if }r\leq R\textrm{~and 0 otherwise}.
\ee
For the relaxation simulations, we have have looked at the region $[-5,5]\times[-5,5]\times[-0.5,0.5]$. We have used a grid of $300\times300\times 30$ points 
covering that domain (the points belonging to that grid have positions $(-5+j\frac{10}{300}-\frac{10}{600},-5+k\frac{10}{300}-\frac{10}{600},-0.5+l\frac{1}{30}-\frac{1}{60})$ 
with $j,k\in\{1,2,\ldots,300\}$ and  $l\in\{1,\ldots,30\}$). 
The domain is divided in $10\times10$ non-overlapping coarse-graining cells (each coarse-graining cell contains $2700$ points). 
For the plots, we have used $46\times 46$ overlapping coarse-graining cells. 
\subsection{Discussion}
\begin{figure}
\centering
\includegraphics[width=\textwidth]{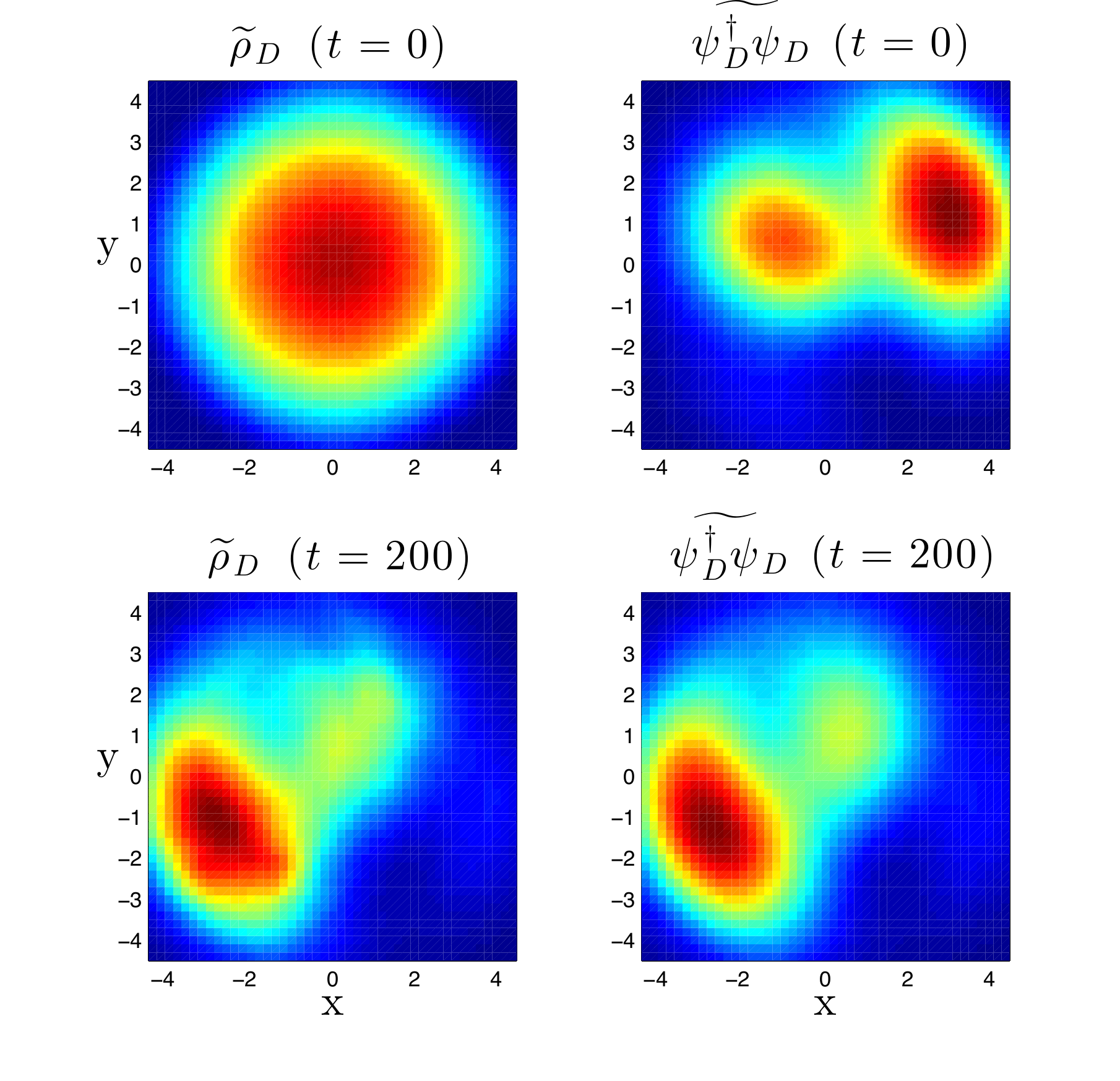}
\caption{\label{fig12}Evolution of $\widetilde\rho_{D}$ and $\widetilde{\psi^\dagger_D\psi_D}$. The initial non-equilibrium distribution is defined 
by $\rho_D(t=0,r,\theta,\phi)=\rho_2(r,\theta,\phi)$, where $\rho_2$ is defined in (\ref{rho03d}). The Dirac spinor is defined in Table \ref{spinor3d}.}
\end{figure}

\begin{figure}
\centering
\includegraphics[width=\textwidth]{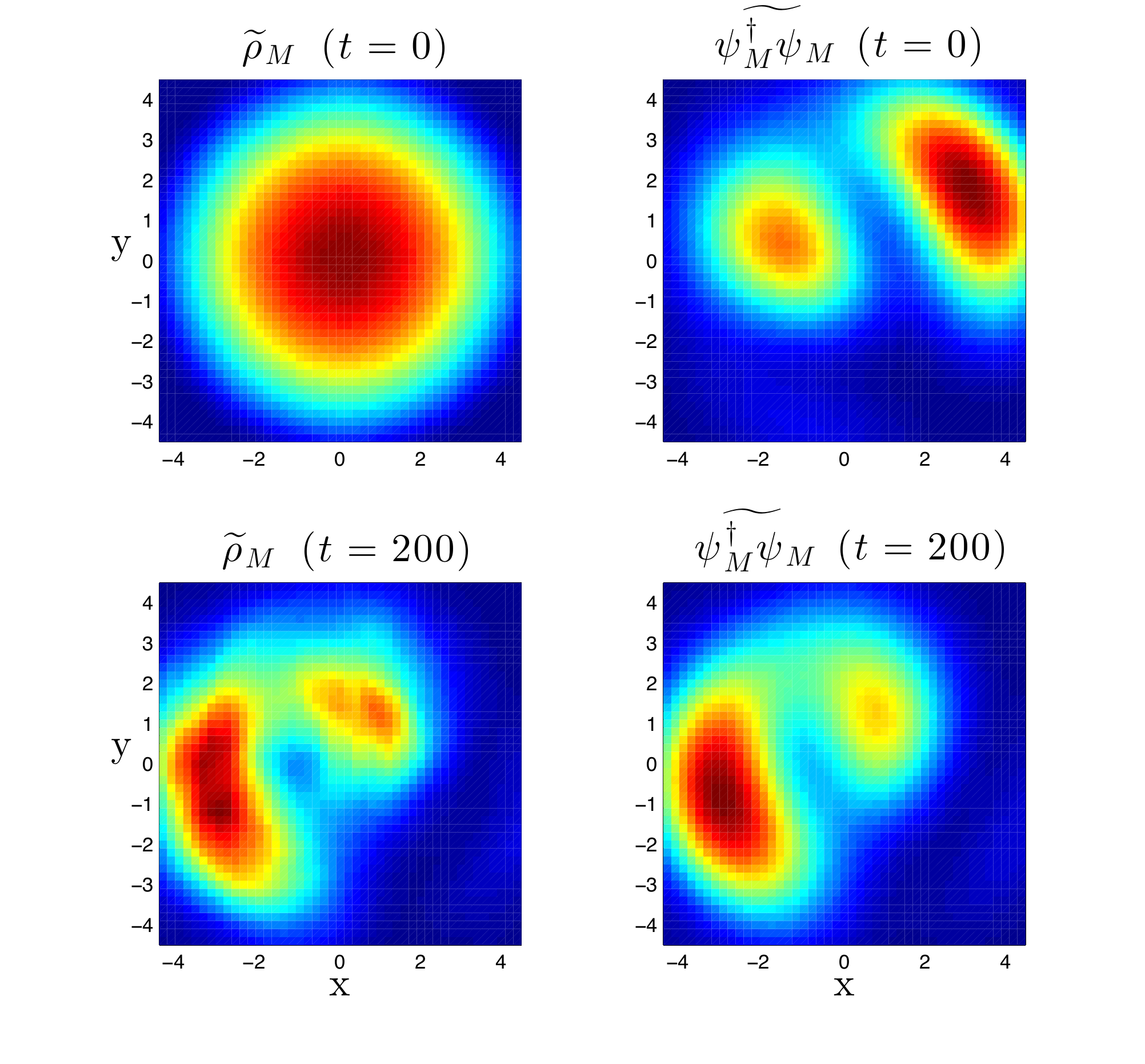}
\caption{\label{fig13}Evolution of $\widetilde\rho_{M}$ and $\widetilde{\psi^\dagger_M\psi_M}$. The initial non-equilibrium distribution is defined 
by $\rho_M(t=0,r,\theta,\phi)=\rho_2(r,\theta,\phi)$, where $\rho_1$ is defined in (\ref{rho02d}). 
The Majorana spinor is obtained from the Dirac spinor defined in Table \ref{spinor3d} thanks to the construction defined at (\ref{addcc}).}
\end{figure}

The results of the simulations can be found in Figures \ref{fig12} and \ref{fig13}, from which we see that the relaxation for the Majorana 
case is only slightly retarded with respect to the one for the Dirac case.
This can either be attributed to the fact that the helix is too loose in the chosen 3D system - to compare examples of trajectories for the 2D and the 3D systems, one can look at Figures \ref{fig10} and \ref{fig11} - or to the difference in the spatial dimension.

In order to settle this question, we need to find a system with a strong helix in 3D. If relaxation is more retarded in this new system (than it is in the old 3D system), then we can 
assume that the spatial dimension is not at the origin of the faster relaxation than we have seen in Figures \ref{fig13}. 
If we look at Tables \ref{spinor2d} and \ref{spinor3d}, we see that the mean energy for the Dirac spinor is closer to the mass in the 2D-system than it is in 
the 3D-system and that the relative energy spread (with respect to the mass) 
is also smaller (and the Majorana spinor is constructed from the Dirac spinor by adding it to its charge conjugate). 
This may explain why the 2D helix is stronger:  the position configuration has to move more slowly on a coarse-grained 
level but the motion is always luminal, which can only be achieved by following a stronger helical trajectory. 
We leave this for future research.

\section{Sub-Compton quantum non-equilibrium}
In this section, we investigate whether quantum non-equilibrium can 
be preserved at the sub-Compton scale in Majorana systems.
We first give a definition of that new notion: a distribution $\rho$ would  be in quantum non-equilibrium at the sub-Compton scale if 
$\bar{\rho}$ present differences with respect to $\overline{\psi^\dagger\psi}$ if and only if the coarse-graining length is much smaller than the Compton wavelength.

In order to find out, we have chosen a small square cell centred at position $(0,3)$ in the 2D system 
($(0,3)$ is just a point such that $\overline{\psi^\dagger\psi}$ is not so different from $\bar\rho$ at the initial time). 
The length of an edge of the square is $0.1$, which is smaller than the Compton wavelength $0.5$. 
We plan to compute $\overline{\psi^\dagger\psi}$ and $\bar\rho$ for various successive times and to study the evolution of the absolute difference between 
the coarse-grained density and the coarse-grained quantum-equilibrium density.
In order to compute the coarse-grained density, we have used $33\times 33$ points distributed uniformly inside the cell. 
We also did a similar simulation with $25\times 25$ points in order to verify that the results obtained in both simulations were not too much different.
The mean difference between the two simulations is $6\%$ but the difference is more important for the nine last cells for which the percentage differences are: 
$(16\%,14\%,3\%,13\%,6\%,11\%,16\%,4\%,9\%)$. For comparison, we also did the Dirac case, for which we also did two simulations. 
The mean percentage difference is $3\%$ and for the nine last cells we have $(8\%,8\%,7\%,14\%,3\%,1\%,1\%,1\%,5\%)$.

\begin{figure}
\centering
\includegraphics[width=\textwidth]{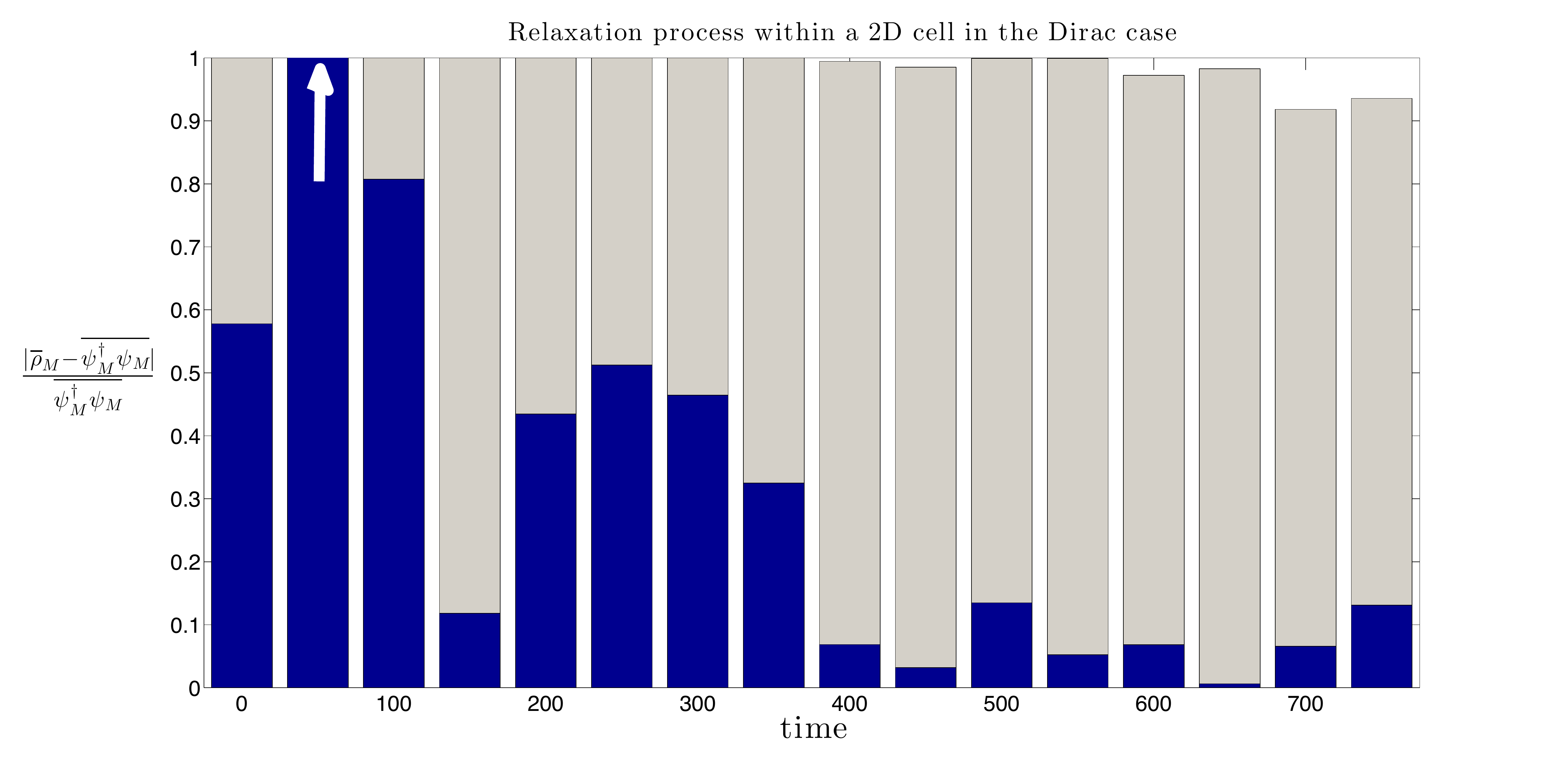}
\caption{\label{fig14}Time-evolution of the difference between $\bar{\rho}_D$ and $\overline{\psi^\dagger_D\psi_D}$ within a cell of surface $(0.1)^2$ centred 
at $(0,3)$. The spinor and the initial non-equilibrium distribution are the ones used in Figure \ref{fig7}.
The gray bars indicate the percentage of good trajectories while the white arrow indicates that the difference exceeds $100\%$.}
\includegraphics[width=\textwidth]{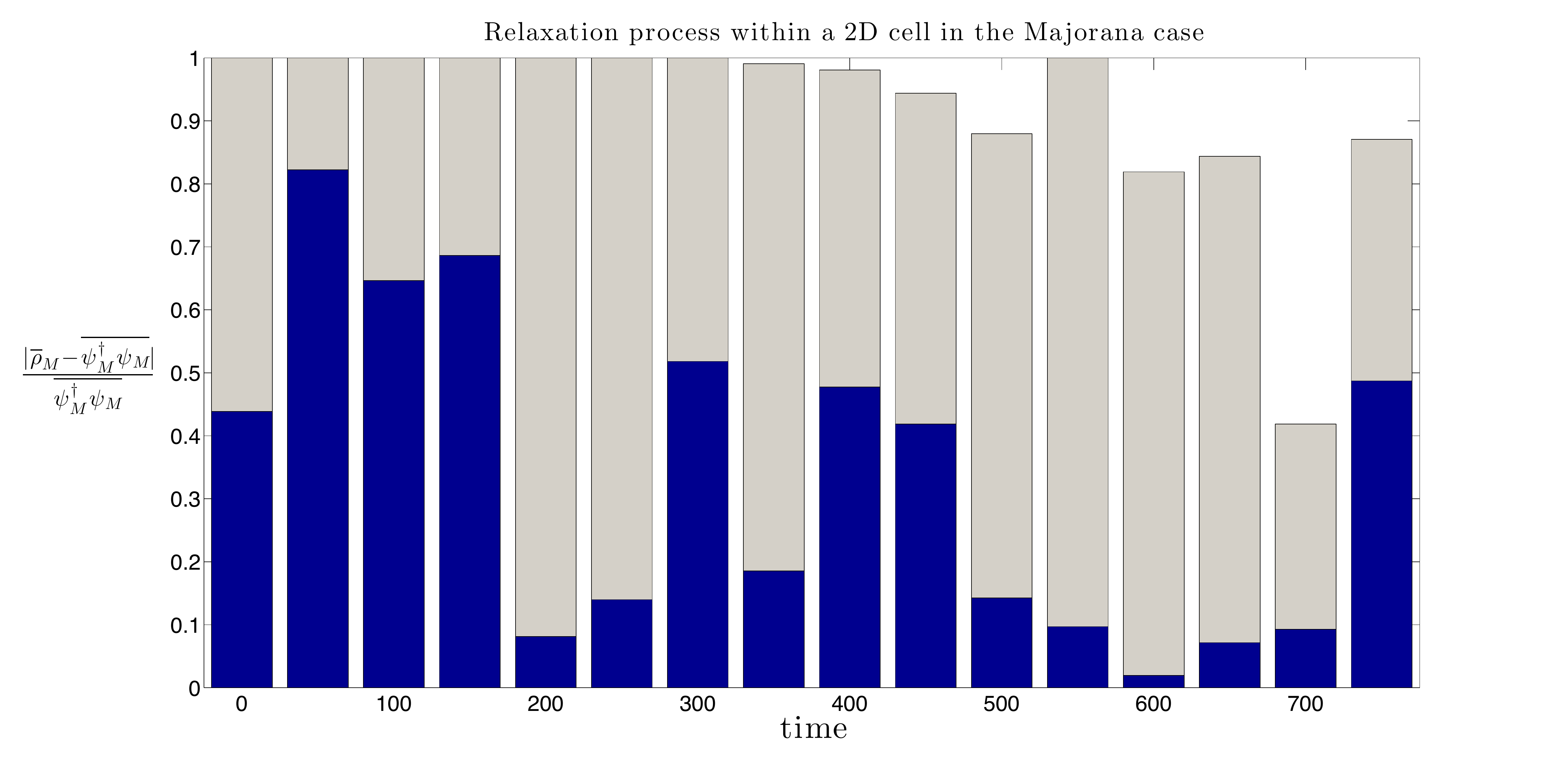}
\caption{\label{fig15}Time-evolution of the difference between $\bar{\rho}_M$ and $\overline{\psi^\dagger_M\psi_M}$ within a cell of surface $(0.1)^2$ centred 
at $(0,3)$. The spinor and the initial non-equilibrium distribution are the ones used in Figure \ref{fig8}.}
\end{figure}

The results for $33\times 33$ points per cell are plotted in Figure \ref{fig14} and Figure \ref{fig15}. 
$\bar{\rho}$ tends to converge to $\overline{\psi^\dagger\psi}$. Howevere there is a retardation in the Majorana case and the amplitudes of the fluctuations are larger. 
Actually the fluctuation at $t=750$ is rather large. To check that it is not a fluke, we have computed the absolute difference between the two densities at $t=760$, 
which is equal to $0.63$ with a percentage of good trajectories equal to $95\%$.
We do the same exercise in the 3D case: we take a cubic cell of centred at $(0,3,0)$ (the length of an edge is still $0.1$ but now it corresponds 
to one tenth of the inside Compton wavelength) and we look at how the distributions differ in the Dirac and the Majorana cases. 
The results go in the same direction but they are not as `spectacular' as those of the  2D-case.
\begin{figure}
\centering
\includegraphics[width=\textwidth]{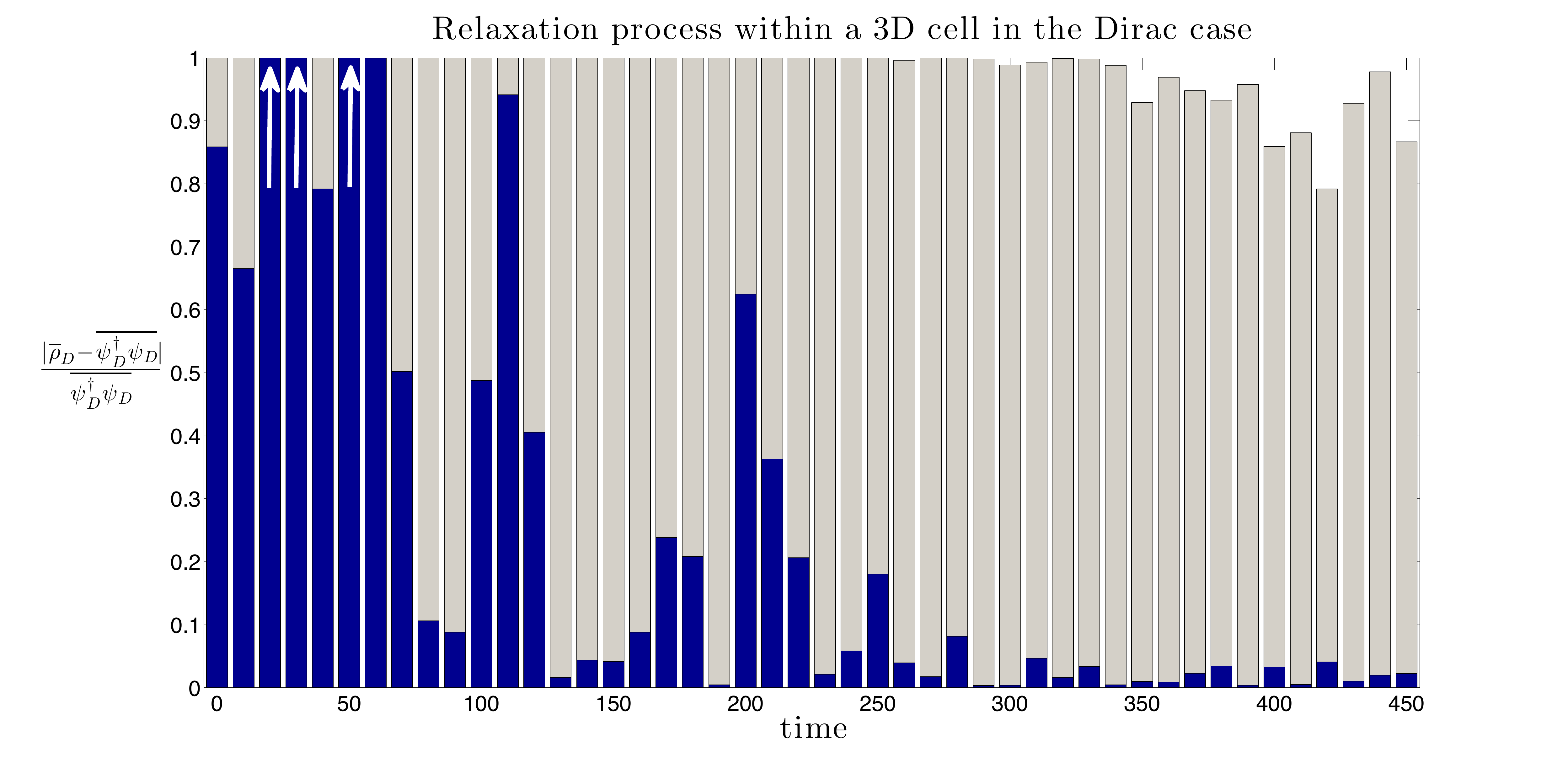}
\caption{\label{fig16}Time-evolution of the difference between $\bar{\rho}_D$ and $\overline{\psi^\dagger_D\psi_D}$ within a cell of volume $(0.1)^3$ centred 
at $(0,3,0)$. The spinor and the initial non-equilibrium distribution are the ones used in Figure \ref{fig12}.
The gray bars indicate the percentage of good trajectories while the white arrows indicate that the difference exceeds $100\%$.}
\includegraphics[width=\textwidth]{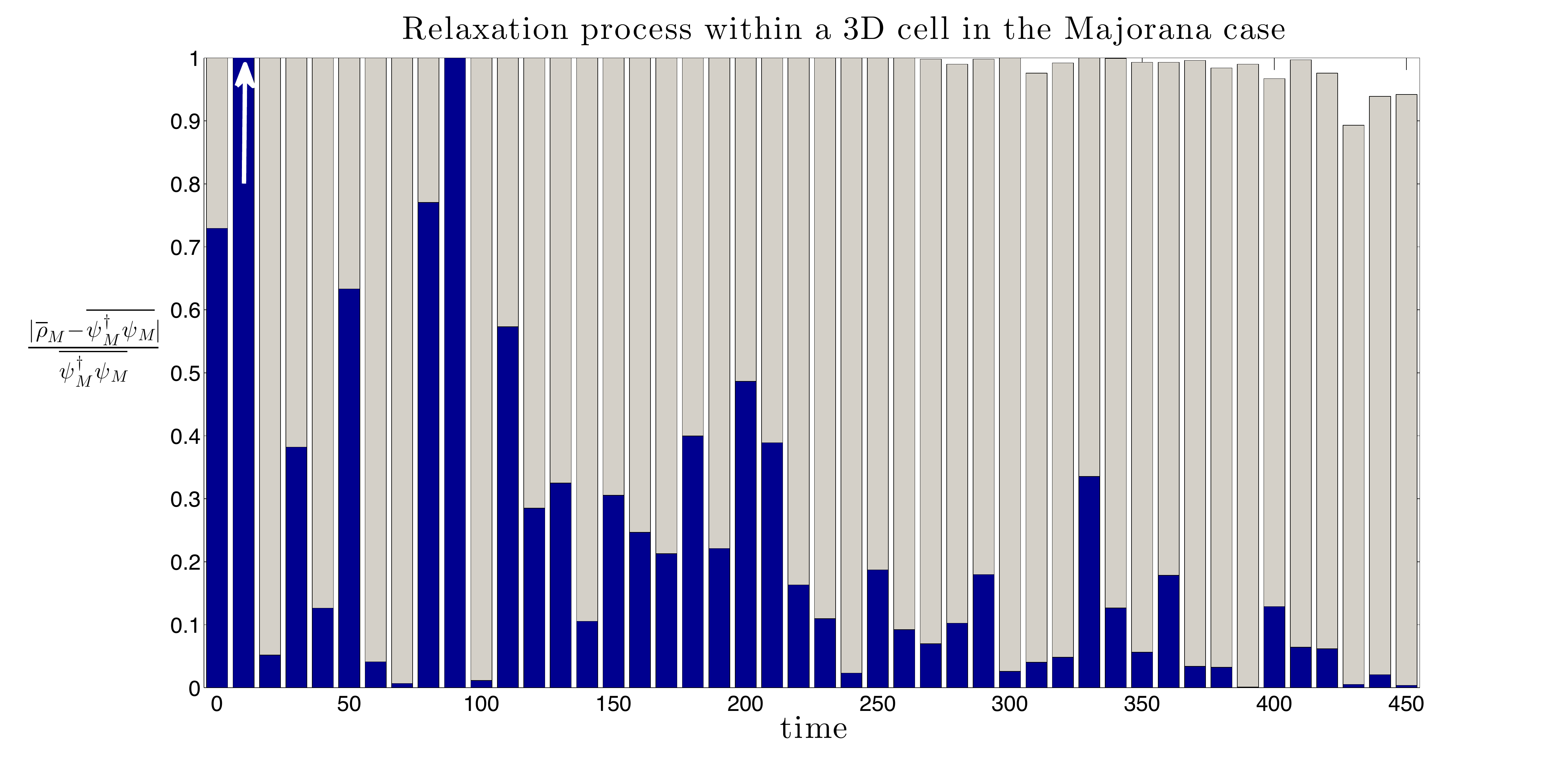}
\caption{\label{fig17}Time-evolution of the difference between $\bar{\rho}_M$ and $\overline{\psi^\dagger_M\psi_M}$ within a cell of volume $(0.1)^3$ centred 
at $(0,3,0)$. The spinor and the initial non-equilibrium distribution are the ones used in Figure \ref{fig13}.
The gray bars indicate the percentage of good trajectories while the white arrow indicates that the difference exceeds $100\%$.}
\end{figure}
\section{Conclusion}
We have built the pilot-wave theory for the Majorana equation. 
We have shown that the trajectories predicted by the theory are always luminal and that they can be strongly helical. 
In the latter case, the diameter of the helix seems to be equal to the Compton wavelength and the trajectory 
appears sub-luminal on a coarse-grained level. We have also done some simulations of the time-evolution of non-equilibrium distributions for quantum systems described by Dirac and Majorana spinors, 
in two and three spatial dimensions. These simulations illustrate that non-equilibrium distributions for Majorana systems, although they relax to 
the quantum equilibrium distribution, are retarded with respect to the ones guided by Dirac spinors 
(this is most visible in the two-dimensional system which exhibits a stronger helix than the three-dimensional one). We have also studied 
the time-evolution of the difference between the distribution $\rho$ and the one predicted by standard quantum theory 
within a small square (and cubic) cell whose length is smaller than the Compton wavelength. 
These simulations also show that the relaxation is retarded but the conclusion that quantum non-equilibrium is preserved at the sub-Compton 
scale cannot be drawn from them at this stage. 

Majorana trajectories are similar to the trajectories obtained from a non-relativistic (spinless) system 
in the neighbourhood of the nodes of the wave-function (which are the only sources of vorticity). In the Majorana case, vortices are 
presumably distributed with a density approximately equal to one vortex per cell of surface equal to the Compton wavelength squared.
Identifying the origin of these Majorana vortices, around which the particles tend to circulate, would give a better understanding of the retardation.
Furthermore, if it possible to increase the vorticity somehow, one can imagine that the retardation process could be reinforced.

If we assume that there was quantum non-equilibrium in the early universe and that the relaxation towards quantum equilibrium in Majorana systems
is retarded (or that quantum non-equilibrium is preserved at sub-Compton scales), one still needs to identify the physical 
systems that could lead to testable predictions. The possible existence of Majorana particles in the early universe (whether neutrinos or supersymmetric partners) 
is not disputed. The main problem comes from the difference between the Majorana spinor and a Majorana particle.
At first, Majorana spinors seem irrelevant for Majorana particles in the full-fledged quantum field theory. 
The reason is that the amplitude describing a Majorana particle $\langle 0|\widehat{\psi}_M(\vec{x})|\Psi_t\rangle$ (where $|0\rangle$ is the free vacuum, 
$\widehat{\psi}_M(\vec{x})$ a Majorana quantum field and $|\Psi_t\rangle$ a state containing one Majorana particle) 
is not invariant under charge conjugation.  Indeed, the annihilation part of the field does not contribute to the amplitude, which is therefore equal to a Dirac spinor. 
However this is not true anymore if one considers the physical vacuum $|V\rangle$ instead of $|0\rangle$, which is rather standard.
Also, the Majorana fermion has three different propagators (instead of one for the Dirac fermion), which does not seem to coincide with the fact 
that it is described by a Dirac spinor.
These observations are meant to indicate that these questions are not settled and deserve further investigation.

Actually the Majorana spinor might even be relevant for Dirac particles. Indeed a Dirac spinor can always be written as the sum of two Majorana spinors. 
In the pilot-wave theory, such a decomposition would lead to a model in which the Dirac electron is made of two alternating particles. 
It was suggested in \cite{struyve_zz} and it was inspired by the a similar zig-zag pilot-wave model in which the Dirac electron is made of two Weyl particles \cite{cowi}.
It was also pointed out in \cite{struyve_zz} that such a decomposition along two Majorana spinors breaks the gauge invariance at the level of the guidance equation. 
However the Lorentz invariance is already broken at the level of the guidance equation and we don't think that it is such an obstacle.
From the pilot-wave theory for the Majorana equation, it appears that such a model would be based on two helices, and an explicit construction 
is also worth doing as future research.

If all this is a dead end, there remains analog systems, like the ones studied in quantum information and condensed matter.
\section*{Acknowledgements}
I acknowledge financial support from a Perimeter Institute Australia Foundations postdoctoral research fellowship. 
This work was also supported by the Australian Research Council Discovery Project DP0880657, ``Decoherence, Time-Asymmetry and the Bohmian View on the 
Quantum World''.
Research at Perimeter Institute is funded by the Government of Canada through Industry Canada and by the Province of Ontario through the Ministry of Research and Innovation.
This research was initiated at Perimeter Institute and Clemson University. The visit at Clemson University, hosted by Antony Valentini, was funded by 
the FQXi mini-grant "Quantum Foundations at Clemson University". 
Some relaxation simulations were performed on the \textit{Griffith University HPC Cluster - V20z}.
I also acknowledge financial support from a Templeton grant.

\end{document}